\documentclass[%
 aip,
 amsmath,amssymb,
 reprint,%
]{revtex4-1}

\usepackage{graphicx}
\usepackage{dcolumn}
\usepackage{bm}

\usepackage[utf8]{inputenc}
\usepackage[T1]{fontenc}
\usepackage{mathptmx}
\usepackage{etoolbox}
\usepackage{subcaption}
\usepackage{xcolor}
\usepackage{url}
\usepackage{hyperref}

\makeatletter
\def\@email#1#2{%
 \endgroup
 \patchcmd{\titleblock@produce}
  {\frontmatter@RRAPformat}
  {\frontmatter@RRAPformat{\produce@RRAP{*#1\href{mailto:#2}{#2}}}\frontmatter@RRAPformat}
  {}{}
}%
\makeatother
\begin{document}

\preprint{AIP/123-QED}

\title
[Extending the mean first-passage-time formalism to reconstructing high barriers]
{Extending the mean first-passage-time formalism to the reconstruction of high free-energy barriers: A top-start approach}

\author{Ying Jiang}
\author{Ivan Saika-Voivod}%
\email{saika@mun.ca}
\affiliation{ 
Department of Physics and Physical Oceanography, Memorial University of Newfoundland, St. John's, NL, Canada, A1B 3X7
}%


\date{\today}

\begin{abstract}

Rare events, such as crystal nucleation in mildly supercooled liquids, are characterized by waiting times that can be much longer than the timescale of the underlying microscopic dynamics. The mean first-passage time (MFPT) formalism [J. Wedekind and D. Reguera, J. Phys. Chem. B 112, 11060 (2008)] provides a route to reconstruct one-dimensional free-energy profiles from the MFPT and steady-state distribution measured along a reaction coordinate, but conventional implementations become inefficient for sufficiently high barriers because spontaneous crossings are rarely observed. We introduce a ``top-start'' extension that initiates trajectories near the barrier top and reconstructs the free-energy profile from their subsequent evolution away from the barrier toward lower-free-energy regions. We establish the equivalence between the top-start and two-region constructions using a Markov-chain transition-matrix formulation. We validate the method using one-dimensional random-walk simulations across low, medium, and high free-energy barriers, with transition-matrix calculations providing an exact benchmark for the corresponding discretized models. The top-start reconstruction remains accurate in the high-barrier regime where conventional MFPT sampling becomes prohibitively inefficient.

\end{abstract}

\maketitle

\section{\label{sec:intro}Introduction}

Many activated processes involve transitions between stable or metastable states that require overcoming a high free-energy barrier. Because such transitions occur on timescales far longer than those of the underlying microscopic dynamics, they are difficult to observe in simulations and are classified as rare events~\cite{frenkel2023understanding}. Despite their infrequency, rare events have important consequences in phenomena ranging from chemical reactions and first-order phase transitions to protein conformational changes~\cite{hanggi_reaction-rate_1990,frenkel2023understanding,protein_conform_Juraszek2012}. However, direct simulations using conventional molecular dynamics (MD) or Monte Carlo (MC) methods~\cite{frenkel2023understanding, allen2017_computer_simulation_liquids} tend to become trapped in metastable basins and rarely sample the barrier region within feasible computational times~\cite{chipot2007_free_energy_calculations, MC1977_inadequet}. This limited exploration of phase space near the top of the barrier complicates both the estimation of transition rates~\cite{hanggi_reaction-rate_1990} and the reconstruction of free-energy landscapes.

Quantitative analysis of activated processes relies on knowledge of the underlying free-energy landscape, which governs both the relative stability of (meta)stable states and the kinetics of transitions between them. Accurate reconstruction of free-energy landscapes requires sufficient sampling near the top of the activation barrier and across multiple regions of phase space. To address the resulting challenges posed by rare-event sampling, a variety of computational methods have been developed. Among the most widely used techniques, umbrella sampling (US) employs a biasing potential on the system to enhance sampling across free-energy barriers and reconstruct the free-energy profile from biased histograms~\cite{TORRIE1974578, TORRIE1977187, Johannes2011_US_review}. While robust, US requires an appropriate predefined reaction coordinate~\cite{coupled_order_parameter1985} and can be computationally demanding. Trajectory-based approaches, such as transition path sampling (TPS) and transition interface sampling (TIS), instead focus on ensembles of reactive trajectories without directly biasing configuration space~\cite{Dellago1998_TPS,TPS2002_review,van2003_TIS}. TPS does not require a predefined reaction coordinate for generating transition paths, while TIS introduces interfaces defined by an order parameter that need not coincide with the true reaction coordinate. Such trajectory-based approaches are therefore well suited to complex processes involving high-dimensional or poorly understood transition mechanisms; however, reconstruction of complete free-energy landscapes typically requires additional procedures~\cite{TPS2019_virtual_interface, Daniele2005_TIS_free_energy, Rogal2011_reweight_TIS}. Forward-flux sampling (FFS) further extends these ideas by integrating the dynamics forward in time only,  allowing it to be applied to nonequilibrium systems that do not satisfy detailed balance~\cite{Allen2005_FFS_first, Allen2006_FFS, Allen2009_FFS_review}.
Although FFS is primarily designed to compute transition rates, extensions have been proposed to extract free-energy information from FFS simulations~\cite{Valeriani2007_FFS_backward_forward, Borrero2009_FFS_US, Vikram2015_FFS_MFPT}.

The mean first-passage time (MFPT) formalism developed by Wedekind and Reguera~\cite{Wedekind2008_kinectic_reconst,Wedekind2007_MFPT} provides a direct connection between the dynamical evolution of a system and its underlying one-dimensional free-energy landscape or {\it profile}. For a reaction coordinate $x$, the MFPT $\tau(x)$ is defined as the average time required for a trajectory starting from a specified initial point $x_0$ to reach $x$ for the first time. Together with the steady-state distribution $P_{\mathrm{st}}(x)$, obtained from the normalized histogram of visited states along the reaction coordinate, the MFPT can be used to reconstruct the corresponding free-energy profile.

The reconstruction can be expressed in terms of two key equations~\cite{Wedekind2008_kinectic_reconst,wedekind_crossover_2009}. The first defines an auxiliary function $B(x)$ on a reaction coordinate $x$,
\begin{equation}
B(x) = - \frac{1}{P_{\mathrm{st}}(x)} \left[ \int_x^{b} P_{\mathrm{st}}(x') \, dx' + \frac{\tau(x) - \tau(b)}{\tau (b)}  \right],
\end{equation}
where $b$ denotes the absorbing boundary at which a trajectory terminates upon arrival. The opposite end of the domain, $x=a$, is taken to be reflecting, so that no probability flux leaves the domain through that boundary. Thus, the reconstruction requires only $\tau(x)$ and $P_{\mathrm{st}}(x)$ measured along the reaction coordinate.

The second equation directly yields the free-energy profile $\beta U(x)$ from $B(x)$:
\begin{equation}
\beta U(x) = \beta U(x_R) + \ln \left( \frac{B(x)}{B(x_R)} \right) - \int_{x_R}^{x} \frac{1}{B(x')} \, dx',
{\label{eq:intro_reconstruct_B_old}}
\end{equation}
where $x_R$ is an interior reference point within the domain. The theoretical basis of these relations is discussed in Section~\ref{subsec:kinetic_reconstruct_method}.

The MFPT formalism provides a bias-free route to reconstructing one-dimensional free-energy profiles from out-of-equilibrium trajectories~\cite{Wedekind2008_kinectic_reconst,hard_spheres_Richard2018} and has been applied to nucleation and crystallization problems~\cite{Sarah2009_mfptCNT,wedekind_crossover_2009}. Its main limitation, which motivates the present work, lies in its reliance on observing spontaneous barrier crossings. For activated barrier-crossing processes, the characteristic escape time grows approximately exponentially with the free-energy barrier height, so spontaneous barrier crossings become increasingly rare as the barrier increases~\cite{hanggi_reaction-rate_1990}. As a result, conventional MFPT sampling becomes inefficient for sufficiently high barriers because adequate statistics near the barrier region require prohibitively long simulations~\cite{Mukhtyar2022,Goswami2023}.

In this paper, we introduce an extension of the MFPT framework, termed the ``top-start'' scheme, designed to overcome this high-barrier sampling limitation. Rather than waiting for trajectories initialized in a low-free-energy region to reach the barrier, top-start trajectories begin near the barrier top, and their subsequent evolution away from the barrier is used to construct the MFPT and steady-state statistics required for free-energy reconstruction. We first develop the idea through a two-region construction, in which the two sides of the barrier are treated separately, and then show how the same information can be recovered using a single top-start simulation setup that allows trajectories to explore both sides of the barrier. The connection between the two schemes is established using transition-matrix and Markov-chain arguments. We then test the reconstruction on one-dimensional random-walk models spanning a range of free-energy barrier heights and compare the results with transition-matrix benchmarks.

The remainder of this paper is organized as follows. Section~II
establishes the theoretical framework used throughout the present work. Section~II\,A reviews the one-dimensional random-walk dynamics and its connection to both the Fokker--Planck equation and a discrete Markov-chain description, thereby providing the common notation and theoretical basis for the subsequent reconstruction and
transition-matrix formulations. Section~II\,B introduces the MFPT and steady-state quantities used in the reconstruction and formulates the problem for the two orientations about a common reflecting boundary, which provides the basis for dividing the reaction-coordinate domain into two regions. Section~II\,C summarizes the corresponding free-energy reconstruction, and Section~II\,D develops the transition-matrix formulation used as an exact discrete benchmark.

Section~III introduces the two-region construction and then develops the top-start scheme, including its transition-matrix representation. Section~IV presents random-walk tests of the conventional, two-region, and top-start reconstructions over a range of barrier heights. Section~V summarizes the main conclusions. Appendix~A reviews the finite-Markov-chain results required in the analysis, Appendix~B uses these results to establish the steady-state and transition-matrix relations underlying the two-region and top-start constructions, and Appendix~C illustrates explicitly how a top-start trajectory is decomposed into the region-restricted trajectories used in the reconstruction.

\section{\label{sec:methodology}Methodology}

\subsection{\label{subsec:random_walk_background}The Kramers problem and Fokker-Planck equation}

The Kramers problem concerns the escape of a Brownian particle from a potential well across an activation barrier due to thermal fluctuations~\cite{zwanzig2001nonequilibrium}. Motivated by the random-walk description of Brownian motion~\cite{pathria2021statistical, 2022Fokker}, we consider a discrete Markov process for a particle evolving along a one-dimensional potential landscape $U(x)$. The coordinate $x$ may represent either the particle position or, more generally, a reaction coordinate describing progress through an activated transition. For the Brownian-particle description, $U(x)$ is the potential energy, whereas for a general reaction coordinate it represents a free-energy profile.

We assume Markovian dynamics, such that the transition probability at each step depends only on the current state and not on the preceding history~\cite{kemeny1960finite}. Let $\rho(x, t)$ denote the probability of observing the Brownian particle at position $x$ at time $t$. The spatial step size of the Brownian particle is denoted as $\Delta x$, and the time step is denoted as $\Delta t$. Assuming that $x$ and its neighboring positions ($x+\Delta x$ and $x-\Delta x$) are not located at any absorbing or reflecting boundary, the probability $\rho(x, t+\Delta t)$ evolves according to
\begin{align}
\rho(x,t+\Delta t) &= P(x \rightarrow x)\,\rho(x,t) \nonumber \\
&\quad + P(x+\Delta x \rightarrow x)\,\rho(x+\Delta x,t) \nonumber \\
&\quad + P(x-\Delta x \rightarrow x)\,\rho(x-\Delta x,t),
\label{eq:Markov_property}
\end{align}
where $P(x_1 \rightarrow x_2)$ denotes the transition probability of moving from initial position $x_1$ at time $t$ to target position $x_2$ at time $t + \Delta t$. In this one-dimensional case, both the initial and target positions are restricted to $x$, $x+\Delta x$, and $x-\Delta x$.

Since the total probability of leaving or staying at a given position must be conserved, the transition probabilities satisfy the condition:
\begin{equation}
P(x \rightarrow x) + P(x \rightarrow x+\Delta x) + P(x \rightarrow x-\Delta x) = 1.
\label{eq:probability_sum}
\end{equation}
Combining Eqs.~\eqref{eq:Markov_property} and
\eqref{eq:probability_sum} gives the net change in the probability of finding the particle at position $x$,
\begin{align}
\rho(x, t+\Delta t) - \rho(x,t) &= - P(x \rightarrow x+\Delta x) \rho(x,t) \nonumber \\
&\quad - P(x \rightarrow x-\Delta x) \rho(x,t) \nonumber \\
&\quad + P(x+\Delta x \rightarrow x) \rho(x+\Delta x,t) \nonumber \\
&\quad + P(x-\Delta x \rightarrow x) \rho(x-\Delta x,t).
\label{eq:net_probability}
\end{align}

As the Brownian particle evolves under the influence of a potential energy $U(x)$, the equilibrium probability distribution $\rho_{\mathrm{eq}}(x)$ follows the Boltzmann distribution,
\[
\rho_{\mathrm{eq}}(x) \propto e^{-\beta U(x)},
\]
where $\beta = 1/(k_BT)$, $T$ is the temperature, and $k_B$ is Boltzmann’s constant.  
To construct transition probabilities consistent with the Boltzmann equilibrium distribution, we impose the detailed balance condition,
\begin{equation}
\rho_{\mathrm{eq}}(x) P(x \rightarrow x \pm \Delta x) = \rho_{\mathrm{eq}}(x\pm \Delta x) P(x \pm \Delta x \rightarrow x).
\label{eq:detailed_balance}
\end{equation}
A convenient choice satisfying detailed balance is the Metropolis acceptance criterion~\cite{swendsen2020introduction,krauth2006statistical}. Denote the probability of accepting the move from position $x$ to $x+\Delta x$ as $A(x \rightarrow x+\Delta x)$,
\begin{equation}
A(x \rightarrow x+\Delta x) = \min\left[1, e^{-\beta (U(x+\Delta x) - U(x))}\right].
\label{eq:metropolis}
\end{equation}
At each step, the Brownian particle attempts to move either to the right or to the left with equal probability. The transition probability for a
move from $x$ to $x+\Delta x$ is therefore,
\begin{equation}
P(x\rightarrow x+\Delta x)
=
\frac{1}{2}
A(x\rightarrow x+\Delta x).
\end{equation}
For clarity, we first consider a locally increasing potential, $U(x+\Delta x)>U(x)>U(x-\Delta x)$. In this case, moves to the right are uphill and are accepted according to the Metropolis factor, whereas moves
to the left are accepted with unit probability. The corresponding Metropolis acceptance probabilities are,
\begin{align}
A(x \rightarrow x + \Delta x) &= e^{-\beta (U(x + \Delta x) - U(x))} \nonumber \\
&= 1 - \beta U'(x)\Delta x - \frac{1}{2} \beta U''(x)(\Delta x)^2 \nonumber \\
&\quad + \frac{1}{2} (\beta U'(x))^2 (\Delta x)^2
+ \mathcal{O}(\Delta x^3), \\
A(x + \Delta x \rightarrow x) &= 1,
\end{align}
where the exponential has been expanded in powers of $\Delta x$. The acceptance probabilities for transitions involving $x-\Delta x$ are obtained analogously. After neglecting terms of order $\mathcal{O}(\Delta x^3)$ and higher, substituting these probabilities into Eq.~\eqref{eq:net_probability} gives,
\begin{align}
\rho(x, t + \Delta t) - \rho(x, t)
&= \frac{(\Delta x)^2}{2}
\frac{\partial^2 \rho(x, t)}{\partial x^2} \nonumber \\
&\quad + \frac{(\Delta x)^2}{2}
\beta U'(x)\frac{\partial \rho(x, t)}{\partial x} \nonumber \\
&\quad + \frac{(\Delta x)^2}{2}
\beta U''(x)\rho(x, t).
\label{eq:approximate_net_probability}
\end{align}

For a locally decreasing potential, the roles of the left- and right-moving transitions are interchanged, and expanding to second order in $\Delta x$
yields the same result. Thus, Eq.~\eqref{eq:approximate_net_probability} is independent of the local sign of $U'(x)$. Dividing Eq.~\eqref{eq:approximate_net_probability} by $\Delta t$ and taking the continuum limit $\Delta x,\, \Delta t\rightarrow 0$, while keeping the diffusion coefficient $D_0 = 
(\Delta x)^2/(2\Delta t)$
finite, we obtain the one-dimensional Fokker-Planck equation~\cite{zwanzig2001nonequilibrium} for a constant diffusion coefficient,
\begin{equation}
\frac{\partial \rho(x,t)}{\partial t}
=
D_0 \frac{\partial}{\partial x}
\left[
\frac{\partial \rho(x,t)}{\partial x}
+\beta U'(x)\rho(x,t)
\right].
\label{eq:simple_FPE}
\end{equation}



For more general activated processes in which the effective diffusion coefficient depends on the reaction coordinate $x$, the corresponding
Fokker-Planck equation can be written as,~\cite{Wedekind2008_kinectic_reconst}
\begin{equation}
\frac{\partial \rho(x, t)}{\partial t} = \frac{\partial}{\partial x} \left[ D(x) e^{-\beta U(x)} \frac{\partial}{\partial x} \left( \rho(x, t) e^{\beta U(x)} \right) \right] = -\frac{\partial J(x, t)}{\partial x},
\label{eq:FPE}
\end{equation}
where $D(x)$ is the coordinate-dependent diffusion coefficient, and $J(x,t)$ is the probability flux.

The Fokker–Planck description provides the continuous framework underlying the MFPT reconstruction used below. In particular, the reconstruction relates the free-energy profile to two quantities obtained from the stochastic dynamics: the mean first-passage time and the steady-state distribution. The following subsections introduce these quantities and their relation to the reconstruction equations. Their discrete Markov-chain counterparts are also used later to construct transition-matrix benchmarks.

\subsection{\label{subsec:mfpt_steady_state}
MFPT, steady-state distribution, and Hill relation}

For one-dimensional diffusive dynamics, mean first-passage times can be obtained from the adjoint equation associated with Eq.~\eqref{eq:FPE}, subject to appropriate boundary
conditions~\cite{hanggi_reaction-rate_1990,zwanzig2001nonequilibrium}. 
Following the free-energy reconstruction approach of Wedekind and Reguera~\cite{Wedekind2007_MFPT,2021Mean},  a reflecting boundary is placed at $a$, and trajectories are initialized at $x_0$. We define $\tau(x)$ as the mean time required for a trajectory to reach the target coordinate $x$ for the first time, and it is given by,
\begin{equation}
\tau(x)
=
\int_{x_0}^x dy\,
\frac{e^{\beta U(y)}}{D(y)}
\int_a^y dz\,e^{-\beta U(z)}.
\label{eq:express_MFPT}
\end{equation}

We consider two possible orientations relative to the common reflecting boundary at $a$. Region~A is the interval $[b_1,a]$, with an absorbing boundary at $b_1<a$, whereas region~B is the interval $[a,b_2]$, with an absorbing boundary at $b_2>a$. In both cases trajectories are initialized at $x_0=a$, where the reflecting boundary condition implies
$\tau^\prime(a)=0$.
Eq.~\eqref{eq:express_MFPT} applies to both orientations:
we denote its restriction to $x\in[b_1,a]$ by $\tau_A(x)$ and its restriction to $x\in[a,b_2]$ by $\tau_B(x)$. For region~A, both integrations in Eq.~\eqref{eq:express_MFPT} are taken toward decreasing $x$, so the two reversed integration directions still give a positive MFPT.

\begin{figure}[htbp]
    \centering
    \includegraphics[width=\linewidth]{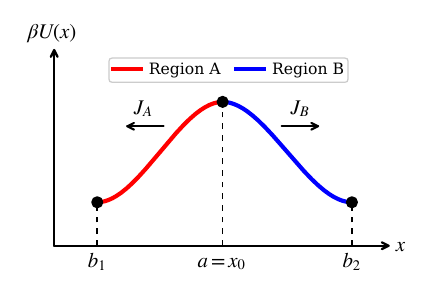}
    \caption[Boundary orientations and steady-state fluxes]{
    Schematic representation of regions~A and B relative to the common reflecting and initial position $a=x_0$. Region~A extends from $a$ toward the absorbing boundary $b_1<a$ and carries a steady-state
    flux $J_A<0$, whereas region~B extends from $a$ toward the absorbing boundary $b_2>a$ and carries a steady-state flux $J_B>0$. The position $a$ need not coincide with the free-energy maximum.
    }
    \label{fig:two_direction_escape}
\end{figure}

To relate the MFPT to the steady-state distribution, we introduce a recycling mechanism~\cite{aristoff2023_WE}: whenever a trajectory reaches the absorbing boundary, it is immediately returned to the initial position $a$. Repeated recycling establishes a nonequilibrium steady state, for which $\partial \rho(x, t)/\partial t=0$. Since $\partial \rho(x, t)/\partial t=-\partial J(x, t)/\partial x$, the probability flux $J$ is constant throughout the region. We denote the corresponding steady-state distribution by $P_{\mathrm{st}}(x)$, and by using Eq.~\eqref{eq:FPE} the steady-state flux is,
\begin{equation}
J
=
-D(x)e^{-\beta U(x)}
\frac{\partial}{\partial x}
\left[
P_{\mathrm{st}}(x)e^{\beta U(x)}
\right].
\label{eq:Jst-Pst}
\end{equation}
Integrating from $x$ to the absorbing boundary $b$ and applying the absorbing boundary condition $P_{\mathrm{st}}(b)=0$ give,
\begin{equation}
P_{\mathrm{st}}(x)
=
J\,e^{-\beta U(x)}
\int_x^b dy\,
\frac{e^{\beta U(y)}}{D(y)}.
\label{eq:Pst}
\end{equation}
This expression applies to either orientation. We denote its restriction to regions~A and B by $P_{\mathrm{st}}^A(x)$ and
$P_{\mathrm{st}}^B(x)$, respectively. For region~B, $b=b_2$ and the probability flux $J_B$ points toward increasing $x$, so $J_B>0$. For region~A, $b=b_1$ and the probability flux $J_A$ points toward decreasing $x$, so $J_A<0$. The two boundary orientations and their corresponding steady-state fluxes are summarized schematically in Fig.~\ref{fig:two_direction_escape}.

For region~B, normalization of the steady-state distribution,
\(
\int_a^{b_2}P_{\mathrm{st}}^B(x)\,dx=1,
\)
gives,
\begin{equation}
\frac{1}{J_B}
=
\int_a^{b_2} dx\,e^{-\beta U(x)}
\int_x^{b_2} dy\,
\frac{e^{\beta U(y)}}{D(y)}.
\label{eq:inverse_JB}
\end{equation}
Changing the order of integration over the same triangular domain yields,
\begin{align}
\frac{1}{J_B}
&=
\int_a^{b_2} dy\,
\frac{e^{\beta U(y)}}{D(y)}
\int_a^y dx\,e^{-\beta U(x)} = \tau_B(b_2).
\label{eq:JB-hill}
\end{align}

For region~A, normalization
\(
\int_{b_1}^{a}P_{\mathrm{st}}^A(x)\,dx=1,
\)
instead gives,
\begin{equation}
\frac{1}{J_A}
=
-\int_{b_1}^{a} dx\,e^{-\beta U(x)}
\int_{b_1}^{x} dy\,
\frac{e^{\beta U(y)}}{D(y)}.
\label{eq:inverse_JA}
\end{equation}
Changing the order of integration gives,
\begin{align}
\frac{1}{J_A}
&=
-\int_{b_1}^{a} dy\,
\frac{e^{\beta U(y)}}{D(y)}
\int_y^a dx\,e^{-\beta U(x)} = -\tau_A(b_1).
\label{eq:JA-hill}
\end{align}
Thus, the MFPT to the absorbing boundary is related to the corresponding steady-state flux by,
\begin{equation}
\tau_A(b_1)=-\frac{1}{J_A},
\qquad
\tau_B(b_2)=\frac{1}{J_B}.
\label{eq:hill_relations}
\end{equation}
These relations are the appropriate forms of the Hill relation~\cite{aristoff2023_WE,torchala2012mean} in our case. The negative sign for region~A reflects the convention that positive flux points toward increasing $x$, whereas the probability flux in region~A is directed toward decreasing $x$. The MFPT and steady-state distribution introduced here are the two dynamical quantities used in the free-energy reconstruction below.

\subsection{\label{subsec:kinetic_reconstruct_method}
Free-energy reconstruction}

Wedekind and Reguera~\cite{Wedekind2008_kinectic_reconst} showed that a one-dimensional free-energy profile can be reconstructed from $\tau(x)$ and $P_{\mathrm{st}}(x)$, both of which can be obtained from simulation. In the notation and coordinate orientation introduced above, the original formulation corresponds
to the region~B. We first summarize this construction and
then obtain the corresponding expression for region~A, where the direction of first passage is reversed. Together, the two reconstructions form the basis of the two-region and top-start schemes introduced below.

For region~B, following Ref.~\cite{Wedekind2008_kinectic_reconst}, we define,
\begin{equation}
A(x)
\equiv
\frac{\partial \tau_B(x)}{\partial x},
\qquad
B(x)
\equiv
A(x)D(x).
\label{eq:AB_regionB}
\end{equation}
The auxiliary function $B(x)$ can be expressed in terms of $P_{\mathrm{st}}^B(x)$ and $\tau_B(x)$ as, 
\begin{equation}
B(x)
=
-\frac{1}{P_{\mathrm{st}}^B(x)}
\left[
\int_x^{b_2}P_{\mathrm{st}}^B(x')\,dx'
-
J_B\left[\tau_B(b_2)-\tau_B(x)\right]
\right].
\label{eq:auxiliary_regionB_flux}
\end{equation}
Using the Hill relation $\tau_B(b_2)=1/J_B$ gives,
\begin{equation}
B(x)
=
-\frac{1}{P_{\mathrm{st}}^B(x)}
\left[
\int_x^{b_2}P_{\mathrm{st}}^B(x')\,dx'
+
\frac{\tau_B(x)-\tau_B(b_2)}
{\tau_B(b_2)}
\right].
\label{eq:auxiliary_regionB}
\end{equation}
Choosing an interior reference position $x_R\in(a,b_2)$, the free-energy profile is reconstructed as,
\begin{equation}
\beta U(x)
=
\beta U(x_R)
+
\ln\left[
\frac{B(x)}{B(x_R)}
\right]
-
\int_{x_R}^{x}\frac{dx'}{B(x')},
\qquad
x\in(a,b_2).
\label{eq:reconstruct_B}
\end{equation}

For region~A, the direction from the reflecting boundary toward the absorbing boundary is reversed. We therefore define
\begin{equation}
A(x)
\equiv
-\frac{\partial \tau_A(x)}{\partial x},
\qquad
B(x)
\equiv
A(x)D(x).
\label{eq:AB_regionA}
\end{equation}
With this definition, $A(x)$ and $B(x)$ remain positive as in the region~B construction. The corresponding auxiliary function is,
\begin{equation}
B(x)
=
-\frac{1}{P_{\mathrm{st}}^A(x)}
\left[
\int_{b_1}^{x}P_{\mathrm{st}}^A(x')\,dx'
+
J_A\left[\tau_A(b_1)-\tau_A(x)\right]
\right].
\label{eq:auxiliary_regionA_flux}
\end{equation}
Using the region~A Hill relation $\tau_A(b_1)=-1/J_A$ gives,
\begin{equation}
B(x)
=
-\frac{1}{P_{\mathrm{st}}^A(x)}
\left[
\int_{b_1}^{x}P_{\mathrm{st}}^A(x')\,dx'
+
\frac{\tau_A(x)-\tau_A(b_1)}
{\tau_A(b_1)}
\right].
\label{eq:auxiliary_regionA}
\end{equation}
Choosing an interior reference position $x_R\in(b_1,a)$, the corresponding free-energy profile is,
\begin{equation}
\beta U(x)
=
\beta U(x_R)
+
\ln\left[
\frac{B(x)}{B(x_R)}
\right]
+
\int_{x_R}^{x}\frac{dx'}{B(x')},
\qquad
x\in(b_1,a).
\label{eq:reconstruct_A}
\end{equation}


The reconstruction equations are evaluated at interior points of the corresponding intervals, with the reference position $x_R$ chosen within the same interior. At the reflecting boundary $a$, $\partial\tau/\partial x=0$ and hence $B(a)=0$, so the logarithmic and integral terms in
Eqs.~\eqref{eq:reconstruct_B} and~\eqref{eq:reconstruct_A} cannot be evaluated by direct substitution. At the absorbing boundaries, $P_{\mathrm{st}}^A(b_1)=0$ and $P_{\mathrm{st}}^B(b_2)=0$, so the corresponding expressions for $B(x)$ are likewise not directly evaluable at $b_1$ and $b_2$. 
Although the reconstruction is carried out at interior points, the boundary values may be understood through the corresponding one-sided limits when needed.
For the two-region reconstruction, each regional free-energy profile is determined up to an additive constant. An overall free-energy shift is physically irrelevant, whereas the relative shift between regions A and B must be determined if the two regional reconstructions are to be combined into a single profile. 
A simple practical matching procedure is to extrapolate the two reconstructed profiles to their common boundary $a$. For instance, one can fit the six interior points nearest $a$ in each region with a quadratic polynomial. 
Evaluating the fitted polynomials at $x=a$ gives estimates $c_0^A$ and $c_0^B$ of the respective one-sided boundary values. The region-B reconstruction can then be shifted by $c_0^A-c_0^B$, yielding a continuous free-energy profile across the two regions without requiring a second independent reference energy.



\subsection{\label{subsec:transfer_matrix_method}Transition-matrix formulation of the Markov chain}

The discrete random-walk description of the Brownian particle introduced in Section~\ref{subsec:random_walk_background} is a discrete-state Markov process, i.e., a Markov chain. For the finite state space considered here, the transition matrix allows direct calculation of the steady-state distribution and MFPT, providing an exact discrete benchmark for the reconstruction methods considered in this work.

We consider a finite Markov chain consisting of $m+2$ states $s_0,s_1,\ldots,s_{m+1}$, where the boundary states $s_0$ and $s_{m+1}$ correspond to the coordinates $x=b_1$ and $x=b_2$, respectively. We denote by $P(s_i\rightarrow s_j)$ the one-step transition probability from state $s_i$ to state $s_j$. The states $s_0$ and $s_{m+1}$ are absorbing, whereas $s_1,\ldots,s_m$ are interior states. The absorbing boundary conditions require $P(s_0\rightarrow s_0)=P(s_{m+1}\rightarrow s_{m+1})=1$. In the first-passage simulations, reaching an absorbing boundary terminates the trajectory; in the corresponding absorbing Markov-chain representation, this termination is represented by retaining the process in the absorbing state with probability one. For an interior state $s_i$, the probability $\rho_i^n$ of observing the process in state $s_i$ at step $n$ satisfies,
\begin{align}
\rho_i^{n+1}
&=P(s_{i-1}\rightarrow s_i)\rho_{i-1}^n
+P(s_i\rightarrow s_i)\rho_i^n \nonumber\\
&\quad+P(s_{i+1}\rightarrow s_i)\rho_{i+1}^n ,
\label{eq:discrete_Markov_chain}
\end{align}
in direct correspondence with Eq.~\eqref{eq:Markov_property}. For compact notation, we define
\begin{equation}
\ell_i=P(s_i\rightarrow s_{i-1}),\quad
r_i=P(s_i\rightarrow s_{i+1}),\quad
h_i=P(s_i\rightarrow s_i),
\end{equation}
so that Eq.~\eqref{eq:discrete_Markov_chain} can equivalently be written as,
\begin{equation}
\rho_i^{n+1}=r_{i-1}\rho_{i-1}^n+h_i\rho_i^n+\ell_{i+1}\rho_{i+1}^n.
\end{equation}
For $1\leq i\leq m$, probability conservation gives $\ell_i+r_i+h_i=1$, consistent with Eq.~\eqref{eq:probability_sum}. At the absorbing boundaries, $r_0=0$ and $\ell_{m+1}=0$. For the Metropolis random walk used here, the left- and right-step probabilities are,
\begin{align}
\ell_i
&=
\frac{1}{2}
\min\!\left[
1,e^{-\beta(U(x_{i-1})-U(x_i))}
\right],
\nonumber\\
r_i
&=
\frac{1}{2}
\min\!\left[
1,e^{-\beta(U(x_{i+1})-U(x_i))}
\right].
\label{eq:metropolis_transition_probabilities_lr}
\end{align}
The complete set of one-step transition probabilities can be assembled into the transition matrix $P$, which characterizes the stochastic evolution of the finite Markov chain.

Here, we adopt the row-vector convention for finite Markov chains~\cite{kemeny1960finite,HUNTER2018100}. Let
$
\boldsymbol{\rho}^{\,n}
=
\left(
\rho_0^n,\rho_1^n,\ldots,\rho_{m+1}^n
\right)
$
denote the probability distribution over all states at step $n$. The matrix element $P_{ij}$ is the probability of a transition from state $s_i$ to state $s_j$, each row of $P$ sums to unity, and the probability distribution evolves according to,
\begin{equation}
\boldsymbol{\rho}^{n+1}=\boldsymbol{\rho}^{n}P.
\label{eq:evolution_matrix}
\end{equation}
The transition matrix for the chain is therefore,
\begin{equation}
P=
\begin{bmatrix}
1      &        &        &        &        &        &        \\
\ell_1    & h_1    & r_1    &        &        &        &        \\
       & \ell_2    & h_2    & r_2    &        &        &        \\
       &        & \ddots & \ddots & \ddots &        &        \\
       &        &        & \ell_{m-1}& h_{m-1}& r_{m-1}&        \\
       &        &        &        & \ell_m    & h_m    & r_m    \\
       &        &        &        &        &        & 1
\end{bmatrix},
\label{eq:AbAb_matrix}
\end{equation}
where entries not shown explicitly are zero.

To reproduce the region-A and region-B constructions introduced in Section~\ref{subsec:kinetic_reconstruct_method}, we select an interior state $s_r$ with $1<r<m$, corresponding to the starting coordinate $x_0=a$, and use it as the reflecting boundary of two regional chains. In region~A, $s_0$ is absorbing and $s_r$ is reflecting. At the reflecting boundary, the probability $r_r$ of an attempted step outside region~A is added to the self-transition probability. The corresponding transition matrix is then,
\begin{equation}
P_A=
\begin{bmatrix}
1      &        &        &        &        &        \\
\ell_1    & h_1    & r_1    &        &        &        \\
       & \ell_2    & h_2    & r_2    &        &        \\
       &        & \ddots & \ddots & \ddots &        \\
       &        &        & \ell_{r-1}& h_{r-1}& r_{r-1}\\
       &        &        &        & \ell_r    & h_r+r_r
\end{bmatrix}.
\label{eq:AbRe_matrix}
\end{equation}

The absorbing process can be converted into a recycling process by reinjecting the process at $s_r$ whenever it reaches $s_0$, following the recycling mechanism introduced in Section~\ref{subsec:mfpt_steady_state}. Replacing the absorbing row by this recycling transition gives,
\begin{equation}
P_A'=
\begin{bmatrix}
       &        &        &        &        & 1      \\
\ell_1    & h_1    & r_1    &        &        &        \\
       & \ell_2    & h_2    & r_2    &        &        \\
       &        & \ddots & \ddots & \ddots &        \\
       &        &        & \ell_{r-1}& h_{r-1}& r_{r-1}\\
       &        &        &        & \ell_r    & h_r+r_r
\end{bmatrix}.
\label{eq:recycle_A_matrix}
\end{equation}
The resulting recycling chain has a steady-state distribution $\boldsymbol{\alpha}_A$ satisfying,
\begin{equation}
\boldsymbol{\alpha}_A P_A'=\boldsymbol{\alpha}_A,\qquad
\sum_i \alpha_{A,i}=1.
\end{equation}
The normalized occupation counts of the transient states before absorption are equivalent to the steady-state distribution obtained from the corresponding recycling process; this relation is proved in Appendix~\ref{sec:appendix_B}. The MFPTs can likewise be obtained from the transition matrix of the recycling chain using Theorem~A.2.3 in Appendix~\ref{sec:appendix_A}.  

For region~B, $s_r$ is reflecting, $s_{m+1}$ is absorbing, and the probability $\ell_r$ of an attempted step outside region~B is added to the self-transition probability at $s_r$, giving,
\begin{equation}
P_B=
\begin{bmatrix}
h_r+\ell_r & r_r       &        &        &        &        \\
\ell_{r+1} & h_{r+1}   & r_{r+1}&        &        &        \\
        & \ddots    & \ddots & \ddots &        &        \\
        &           & \ell_{m-1}& h_{m-1}& r_{m-1}&        \\
        &           &        & \ell_m    & h_m    & r_m    \\
        &           &        &        &        & 1
\end{bmatrix}.
\label{eq:ReAb_matrix}
\end{equation}
Recycling the process from $s_{m+1}$ back to $s_r$ gives,
\begin{equation}
P_B'=
\begin{bmatrix}
h_r+\ell_r & r_r       &        &        &        &        \\
\ell_{r+1} & h_{r+1}   & r_{r+1}&        &        &        \\
        & \ddots    & \ddots & \ddots &        &        \\
        &           & \ell_{m-1}& h_{m-1}& r_{m-1}&        \\
        &           &        & \ell_m    & h_m    & r_m    \\
1       &           &        &        &        &
\end{bmatrix}.
\label{eq:recycle_B_matrix}
\end{equation}

The transition-matrix calculation therefore provides the exact discrete steady-state distributions and MFPTs for the same stochastic dynamics sampled by the random-walk simulations. These exact results calculated by matrices serve as benchmarks for the region-A, region-B, and top-start reconstructions below. Appendix~\ref{sec:appendix_A} summarizes standard results from finite Markov-chain theory used in these calculations, while Appendix~\ref{sec:appendix_B} establishes the connection between the absorbing and recycling descriptions and demonstrates the equivalence between the top-start and two-region constructions.

\section{\label{sec:top_start_scheme}Extending MFPT reconstruction: top-start scheme}

For large free-energy barriers, spontaneous barrier-crossing events become rare, making conventional MFPT reconstruction increasingly inefficient. To overcome this limitation, we consider simulations initialized near the barrier top, where trajectories can relax toward lower-free-energy regions on either side. We first introduce an auxiliary two-region construction that underpins the top-start scheme.

\subsection{\label{subsec:two_region_method}Two-region method}

As established in Section~\ref{subsec:kinetic_reconstruct_method}, the free-energy profile can be reconstructed separately on region~A, $b_1\leq x\leq a$, and region~B, $a\leq x\leq b_2$. The two regions are treated in independent simulations, each initialized at $a$, which also serves as the reflecting boundary in both setups, with absorbing boundaries at $b_1$ and $b_2$ for regions~A and~B, respectively.

At the simulation level, reaching an absorbing boundary terminates the trajectory, whereas a reflecting boundary prevents the trajectory from leaving the prescribed region. For the discrete random-walk simulations considered below, an attempted step across the reflecting boundary is rejected and the walker remains at $a$. More generally, the particular implementation of the reflecting condition depends on the underlying dynamics.

For each region, an ensemble of independent trajectories is used to determine the quantities required by the reconstruction formalism. The first-passage times to positions along the reaction coordinate are averaged over independent trajectories to obtain $\tau_A(x)$ and $\tau_B(x)$. The corresponding steady-state distributions $P_{\mathrm{st}}^A(x)$ and $P_{\mathrm{st}}^B(x)$ are obtained by accumulating the number of visits to each position over the trajectories and normalizing the resulting distributions. These quantities are then used in Eqs.~\eqref{eq:auxiliary_regionA} and~\eqref{eq:reconstruct_A} for region~A and Eqs.~\eqref{eq:auxiliary_regionB} and~\eqref{eq:reconstruct_B} for region~B. The complete free-energy profile is obtained by matching the relative free-energy offset between the two regional reconstructions, as described in Section~\ref{subsec:kinetic_reconstruct_method}. We refer to the overall sampling and reconstruction procedure as the two-region method.

Because trajectories begin near the barrier top rather than waiting for spontaneous barrier-crossing events from a metastable region, the two-region construction avoids the sampling difficulty associated with high barriers. However, it requires two separate simulation setups and the explicit implementation of a reflecting boundary at $a$. These practical limitations motivate the top-start scheme introduced below.

\subsection{\label{subsec:top_start}Top-start scheme}

To eliminate the need for separate simulation setups and their respective reflecting boundaries as required by the two-region method, we introduce the top-start scheme. A simulation is initialized at $a$, chosen near the top of the free-energy barrier, with absorbing boundaries at $b_1$ and $b_2$. In contrast to the two-region construction, $a$ serves only as the starting point and is not a reflecting boundary.

Starting from $a$, the system evolves freely in either direction until it reaches one of the absorbing boundaries, $b_1$ or $b_2$. Following the recycling procedure introduced in Section~\ref{subsec:mfpt_steady_state}, the system is reinjected at $a$ after one of the absorbing boundaries is reached and continues to evolve until the top-start realization terminates. A top-start realization terminates once both absorbing boundaries have been visited; before this occurs, the same absorbing boundary may be reached multiple times. The treatment of these repeated boundary encounters is described below. Each top-start realization therefore produces one full top-start trajectory, extending from its initialization at $a$ to termination. Repeating the procedure generates an ensemble of independent top-start realizations.

Subsequent postprocessing recovers the same region-specific first-passage and steady-state statistics required by the two-region method. The region-specific data are obtained by constructing a region-restricted trajectory for each region from a full top-start trajectory. For region~A, only states in $[b_1,a]$ are retained, while excursions into region~B are omitted; the corresponding region-restricted time advances only along the retained trajectory. Region~B is treated analogously. If $b_1$ is reached before the top-start realization terminates, the current region-A trajectory is completed and, after reinjection at $a$, a new region-A trajectory begins, whereas the current region-B trajectory continues without being reset. The analogous procedure applies if $b_2$ is reached before termination. When the top-start realization terminates, any region-restricted trajectory that has not reached its corresponding absorbing boundary is incomplete and is discarded.

Within a given top-start realization, more than one completed region-restricted trajectory may be obtained for the region whose absorbing boundary is reached first. The other region yields one completed region-restricted trajectory when its absorbing boundary is reached and the realization terminates. For the random-walk simulations reported below, when multiple region-restricted trajectories are completed for the same region within a single top-start realization, their first-passage times (FPTs) and visit counts are averaged position by position before averaging over independent realizations. Thus, each top-start realization contributes one region-specific FPT array and visit-count histogram for each region. The resulting FPT arrays and visit-count histograms are then averaged over top-start realizations to obtain $\tau_A(x)$, $P_{\mathrm{st}}^A(x)$, $\tau_B(x)$, and $P_{\mathrm{st}}^B(x)$, which are used in Eqs.~\eqref{eq:auxiliary_regionA} and~\eqref{eq:reconstruct_A} for region~A and Eqs.~\eqref{eq:auxiliary_regionB} and~\eqref{eq:reconstruct_B} for region~B.

This within-realization averaging is important for proper sampling. If all completed region-restricted trajectories were instead treated as independent samples, a top-start realization that produces more completed trajectories for one region would contribute more heavily to the ensemble average for that region.  Averaging first within each realization ensures that each independent top-start realization provides equal statistical weight for each region. A representative random-walk trajectory decomposition and a step-by-step six-state example are given in Appendix~\ref{app:top_start_RW_example} to illustrate the construction of the region-restricted trajectories, region-restricted time, stopping rule, and averaging procedure.

The top-start scheme therefore obtains the same region-specific inputs required by the two-region reconstruction from a single simulation setup that does not require imposing a reflecting boundary at $a$. The equivalence between the top-start trajectories and the corresponding two-region trajectories is established below using the transition-matrix formulation. 

\subsection{\label{subsec:top_start_transfer_matrix}Transition-matrix validation of the top-start scheme}


The top-start scheme can be related directly to the two-region method within the finite-Markov-chain framework introduced in Section~\ref{subsec:transfer_matrix_method}. Here we give the essential argument; Appendix~\ref{subsec:prove_top_start} provides an explicit matrix derivation.

For the Markov-chain analysis, we consider the recycling chain associated with the top-start scheme. Whenever the process reaches either boundary state, $s_0$ or $s_{m+1}$, it is reinjected at the starting state $s_r$. The transition matrix of this recycling chain is,
\begin{equation}
P'=
\begin{bmatrix}
       &        &        &        & 1      &        &        &        &        \\
\ell_1    & h_1    & r_1    &        &        &        &        &        &        \\
       & \ell_2    & h_2    & r_2    &        &        &        &        &        \\
       &        & \ddots & \ddots & \ddots &        &        &        &        \\
       &        &        & \ell_r    & h_r    & r_r    &        &        &        \\
       &        &        &        & \ddots & \ddots & \ddots &        &        \\
       &        &        &        &        & \ell_{m-1}& h_{m-1}& r_{m-1}&        \\
       &        &        &        &        &        & \ell_m    & h_m    & r_m    \\
       &        &        &        & 1      &        &        &        &
\end{bmatrix},
\label{eq:recycle_matrix}
\end{equation}
where entries not shown explicitly are zero. 
In the associated absorbing process, $s_0$ and $s_{m+1}$ are absorbing boundary states. In the recycling process, reaching either boundary is instead followed by reinjection at $s_r$ with probability $1$.


For region~A, we retain the state set,
\[
\mathcal{S}_A=\{s_0,s_1,\ldots,s_r\},
\]
and omit excursions outside this set in the region-restricted description. Because $s_r$ is the only state connecting $\mathcal{S}_A$ to its complement, the process must both leave and re-enter $\mathcal{S}_A$ through $s_r$. When the excursion outside $\mathcal{S}_A$ is omitted, it is therefore represented by a self-transition $s_r\rightarrow s_r$. The resulting transition rule is identical to the reflecting-boundary rule at $s_r$ used in the region-A recycling matrix $P_A'$ defined in Eq.~\eqref{eq:recycle_A_matrix}, with all other transitions within $\mathcal{S}_A$ remaining unchanged.

The same argument applies to region~B, for which,
\[
\mathcal{S}_B=\{s_r,s_{r+1},\ldots,s_{m+1}\}.
\]
An excursion outside $\mathcal{S}_B$ likewise begins and ends at $s_r$ and therefore reduces to a self-transition at $s_r$ when the excursion is omitted. The resulting transition matrix is consequently identical to the region-B recycling matrix $P_B'$ defined in Eq.~\eqref{eq:recycle_B_matrix}.

Thus, the region-restricted trajectories extracted from the top-start scheme obey the same transition rules as the corresponding two-region trajectories. The associated Markov chains are therefore equivalent and yield identical steady-state distributions and MFPTs. This justifies obtaining $P_{\mathrm{st}}^A(x)$, $\tau_A(x)$, $P_{\mathrm{st}}^B(x)$, and $\tau_B(x)$ from the region-restricted top-start trajectories for use in the corresponding reconstruction equations. 

\section{\label{sec:random_walk_examples}Random-walk validation}

We validate the MFPT reconstruction procedures using 2000 independent MC random walk (RW) simulations evolving in one-dimensional double-well potentials. The MC dynamics follow the Metropolis acceptance criterion and provide a discrete random-walk realization of the Brownian-particle description introduced in Section~\ref{subsec:random_walk_background}. Each walker moves with a spatial step size of $\Delta x=0.01$. The diffusion coefficient is set to $D_0=0.01$, corresponding to a time step of $\Delta t=(\Delta x)^2/(2D_0)=0.005$.

\begin{figure}[htbp]
    \centering
    \includegraphics[
        width=0.95\linewidth
    ]{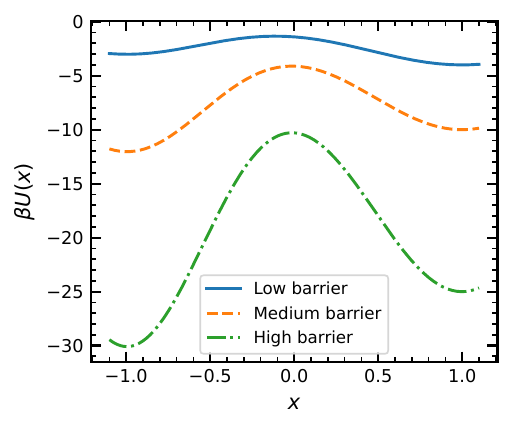}
    \caption{
    Model double-well potentials used in the RW simulations for the low-, medium-, and high-barrier cases. The corresponding barrier heights, measured from the left-well minimum, are approximately $2$, $8$, and $20\,k_{\mathrm B}T$, respectively.
    }
    \label{fig:model_potentials}
\end{figure}

The model potentials are constructed as the sum of two Gaussian wells,
\begin{equation}
\beta U(x)
=
-A_1
\exp\left[
-\frac{(x-\mu_1)^2}{2\sigma_1^2}
\right]
-A_2
\exp\left[
-\frac{(x-\mu_2)^2}{2\sigma_2^2}
\right].
\label{eq:double_gaussian}
\end{equation}
The Gaussian centers and widths are fixed at $\mu_1=-1$, $\mu_2=1$, $\sigma_1=0.5$, and $\sigma_2=0.6$. The amplitudes are set to $(A_1,A_2)=(3,4)$, $(12,10)$, and $(30,25)$ for the low-, medium-, and high-barrier cases, respectively. These landscapes, shown in Fig.~\ref{fig:model_potentials}, provide controlled test cases in which barrier-crossing events become increasingly rare. Because the model potentials are known analytically, they also provide direct benchmarks for the reconstructed free-energy profiles.

For each reconstruction scheme, the steady-state distributions and MFPTs obtained from RW simulations are compared with transition-matrix (TM) calculations. For the corresponding Markov chain, defined using the same spatial discretization and transition probabilities, the TM calculations provide exact results for the discretized model, up to numerical precision. The RW estimates, by contrast, are subject to finite-sampling
uncertainty. 

The calculation of MFPTs and steady-state distributions from the TM is described in Appendix~\ref{subsec:regular_chain_def}. The resulting TM quantities are used in Eqs.~\eqref{eq:auxiliary_regionA} and~\eqref{eq:reconstruct_A} for region~A and Eqs.~\eqref{eq:auxiliary_regionB} and~\eqref{eq:reconstruct_B} for region~B to reconstruct the corresponding free-energy profiles.
The codes used for data generation, analysis, and plotting are publicly available~\cite{MFPT_top_start_github}.

\subsection{Reguera's MFPT reconstruction: low- and medium-barrier cases}

We first apply the conventional Reguera reconstruction described in Section~\ref{subsec:kinetic_reconstruct_method}. A reflecting boundary is placed at $a=x_0=-1.1$ and an absorbing boundary at $b=1.1$. With the model potentials shown in Fig.~\ref{fig:model_potentials}, sampling the region on the right-hand side of the barrier requires trajectories to cross the barrier from left to right. As the barrier height increases, such crossings become increasingly rare. We therefore use the low- and medium-barrier potentials to illustrate both the reconstruction and the onset of this sampling limitation.
For clarity, we note that a RW located at $a$ attempts to move to the left with probability 0.5, but the move is always rejected; a move to the right is attempted with probability 0.5 and is accepted with the usual Metropolis criterion.

\begin{figure*}[t]
    \centering
    \begin{subfigure}[b]{0.235\textwidth}
        \centering
        \includegraphics[width=\linewidth]{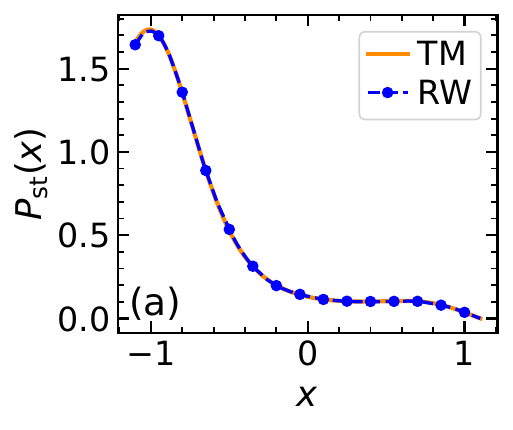}
        \label{fig:reguera_pst_low}
    \end{subfigure}\hfill
    \begin{subfigure}[b]{0.235\textwidth}
        \centering
        \includegraphics[width=\linewidth]{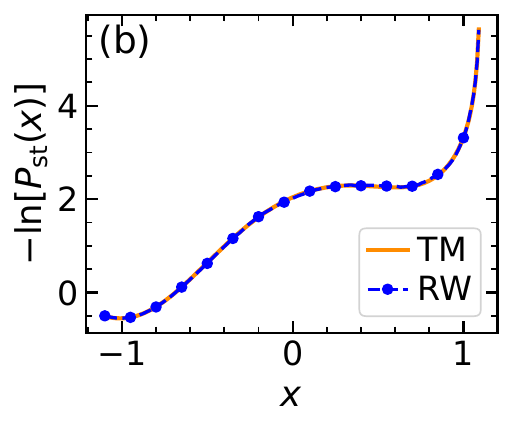}
        \label{fig:reguera_lnpst_low}
    \end{subfigure}\hfill
    \begin{subfigure}[b]{0.235\textwidth}
        \centering
        \includegraphics[width=\linewidth]{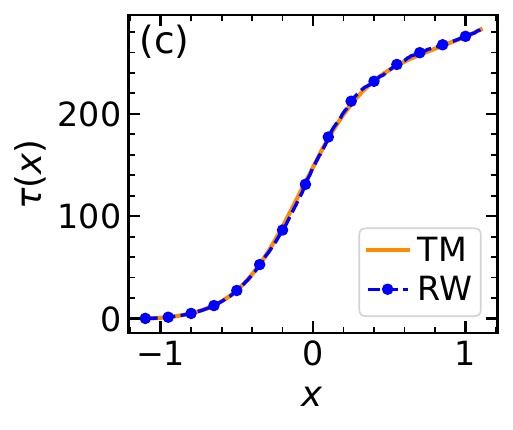}
        \label{fig:reguera_mfpt_low}
    \end{subfigure}\hfill
    \begin{subfigure}[b]{0.235\textwidth}
        \centering
        \includegraphics[width=\linewidth]{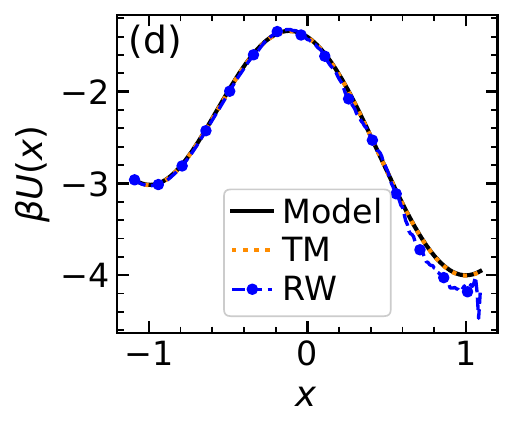}
        \label{fig:reguera_reconst_low}
    \end{subfigure}

    \begin{subfigure}[b]{0.235\textwidth}
        \centering
        \includegraphics[width=\linewidth]{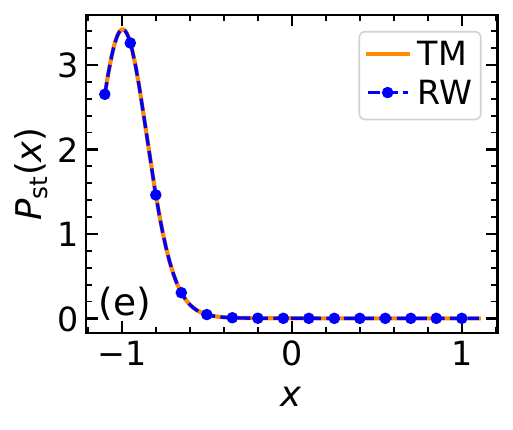}
        \label{fig:reguera_pst_medium}
    \end{subfigure}\hfill
    \begin{subfigure}[b]{0.235\textwidth}
        \centering
        \includegraphics[width=\linewidth]{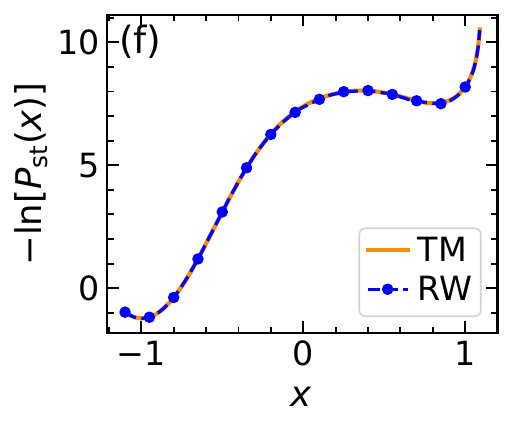}
        \label{fig:reguera_lnpst_medium}
    \end{subfigure}\hfill
    \begin{subfigure}[b]{0.235\textwidth}
        \centering
        \includegraphics[width=\linewidth]{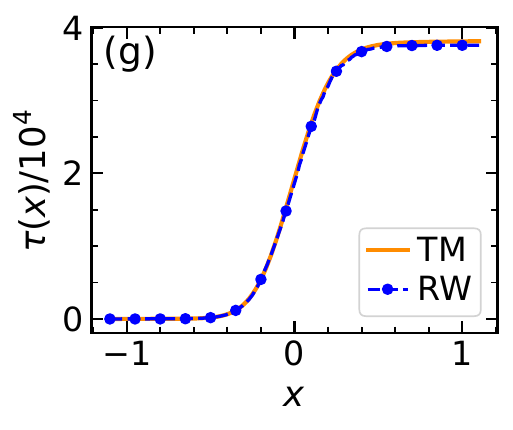}
        \label{fig:reguera_mfpt_medium}
    \end{subfigure}\hfill
    \begin{subfigure}[b]{0.235\textwidth}
        \centering
        \includegraphics[width=\linewidth]{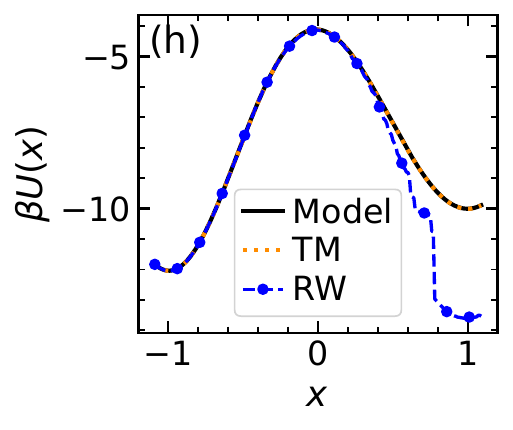}
        \label{fig:reguera_reconst_medium}
    \end{subfigure}

    \caption[Conventional MFPT reconstruction for the low- and medium-barrier cases]
    {Conventional MFPT reconstruction for the low- and medium-barrier model potentials. The first and second rows correspond to the low- and medium-barrier cases, respectively. From left to right, the columns show $P_{\mathrm{st}}(x)$, $-\ln P_{\mathrm{st}}(x)$, the MFPT $\tau(x)$, and the reconstructed free-energy profile $\beta U(x)$. All the RW results are compared with TM calculations, while the reconstructed free-energy profiles are compared with the corresponding model potentials.}
    \label{fig:reguera_panels_combined}
\end{figure*}

Fig.~\ref{fig:reguera_panels_combined} presents the results of using Reguera's method, and serves to validate our RW and TM calculations. The first and second rows correspond to the low- and medium-barrier potentials, respectively. From left to right, the columns show $P_{\mathrm{st}}(x)$, $-\ln P_{\mathrm{st}}(x)$, $\tau(x)$, and the reconstructed free-energy profile $\beta U(x)$. The logarithmic representation is included because the steady-state probability beyond the barrier becomes very small in the medium-barrier case, making its variation difficult to distinguish on a linear scale.

For the low-barrier case, the RW estimates of $P_{\mathrm{st}}(x)$ and $\tau(x)$ closely follow the TM results, and the reconstructed free-energy profile reproduces the original model potential.

For the medium-barrier case, the MFPT exhibits a sigmoidal shape, with a plateau value of order $10^4$, compared with order $10^2$ for the low-barrier case. This increase is consistent with the expected exponential growth of the escape time with barrier height. The RW and TM results remain consistent through the barrier region, while the RW data become progressively noisier on the far side of the barrier near the absorbing boundary.  

The barrier of the high-barrier model potential is approximately $12\,k_{\mathrm B}T$ higher than in the medium-barrier case. Based on the exponential dependence of the escape time on barrier height, this corresponds to an increase of roughly five orders of magnitude in the MFPT, from order $10^4$ to order $10^9$. A separate TM calculation gives an MFPT of order $10^9$, consistent with this estimate; the corresponding RW calculation was therefore not pursued because direct barrier-crossing sampling would be computationally prohibitive. The TM-based reconstruction also shows deviations from the model potential beyond the barrier, indicating numerical sensitivity in the reconstruction procedure.
These results illustrate the increasing difficulty of conventional MFPT sampling as the barrier height increases and motivate the region-specific reconstruction considered next. 

\subsection{Two-region reconstruction}

We next apply the two-region construction of Section~\ref{subsec:two_region_method}. A reflecting boundary is placed at $a=x_0=-0.1$, with absorbing boundaries at $b_1=-1.1$ and $b_2=1.1$. Separate RW simulations are performed in regions~A and~B. The resulting $P_{\mathrm{st}}^A(x)$, $P_{\mathrm{st}}^B(x)$, $\tau_A(x)$, and $\tau_B(x)$ are then used in the corresponding region-specific reconstruction equations. We apply the same procedure to all three model potentials.

\begin{figure*}[t]
    \centering
    \begin{subfigure}[b]{0.32\textwidth}
        \centering
        \includegraphics[width=\linewidth]{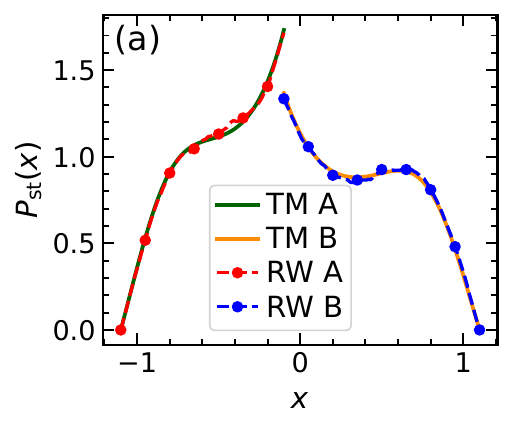}
        \label{fig:two_low_pst}
    \end{subfigure}\hfill
    \begin{subfigure}[b]{0.32\textwidth}
        \centering
        \includegraphics[width=\linewidth]{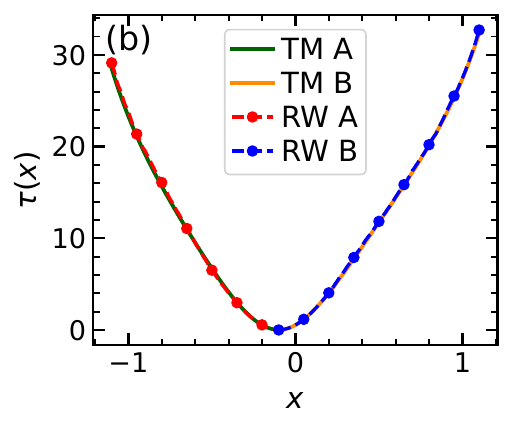}
        \label{fig:two_low_mfpt}
    \end{subfigure}\hfill
    \begin{subfigure}[b]{0.32\textwidth}
        \centering
        \includegraphics[width=\linewidth]{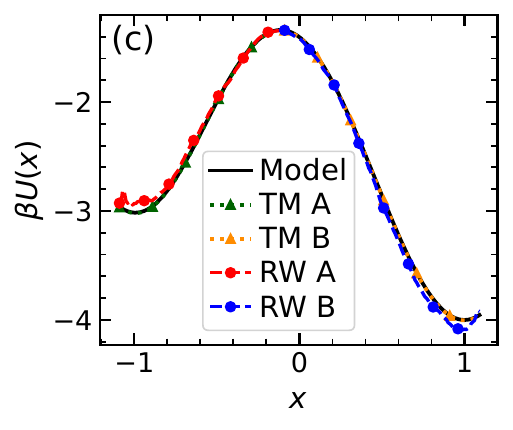}
        \label{fig:two_low_reconst}
    \end{subfigure}

    \begin{subfigure}[b]{0.32\textwidth}
        \centering
        \includegraphics[width=\linewidth]{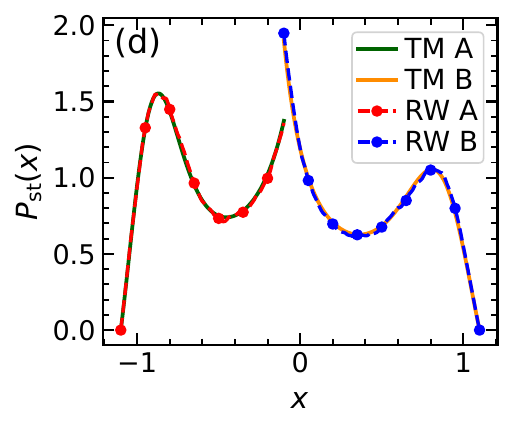}
        \label{fig:two_medium_pst}
    \end{subfigure}\hfill
    \begin{subfigure}[b]{0.32\textwidth}
        \centering
        \includegraphics[width=\linewidth]{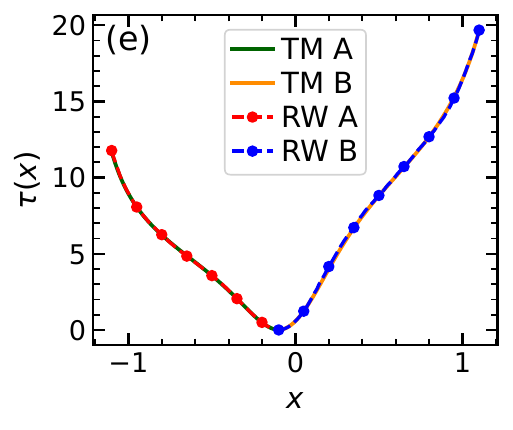}
        \label{fig:two_medium_mfpt}
    \end{subfigure}\hfill
    \begin{subfigure}[b]{0.32\textwidth}
        \centering
        \includegraphics[width=\linewidth]{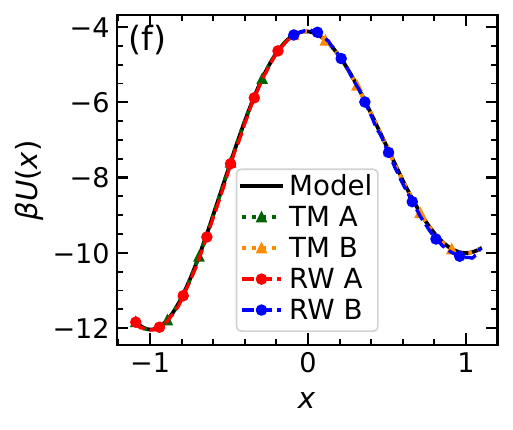}
        \label{fig:two_medium_reconst}
    \end{subfigure}

    \begin{subfigure}[b]{0.32\textwidth}
        \centering
        \includegraphics[width=\linewidth]{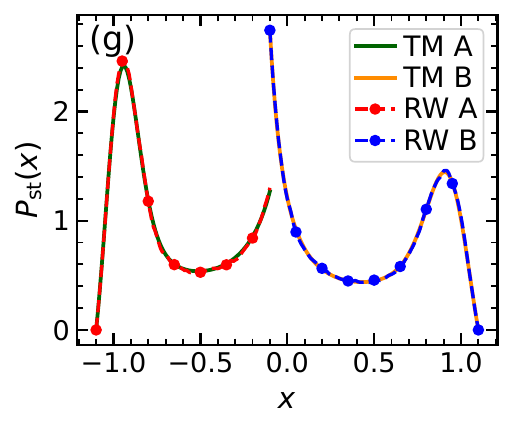}
        \label{fig:two_high_pst}
    \end{subfigure}\hfill
    \begin{subfigure}[b]{0.32\textwidth}
        \centering
        \includegraphics[width=\linewidth]{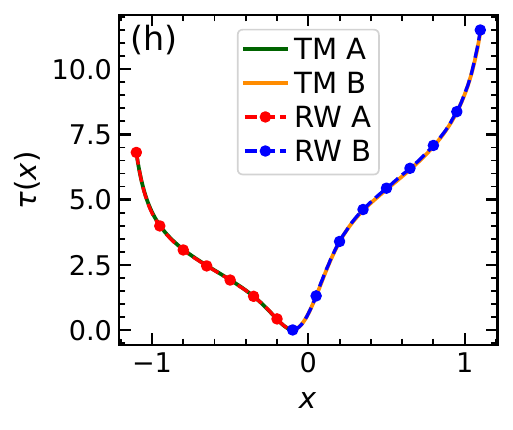}
        \label{fig:two_high_mfpt}
    \end{subfigure}\hfill
    \begin{subfigure}[b]{0.32\textwidth}
        \centering
        \includegraphics[width=\linewidth]{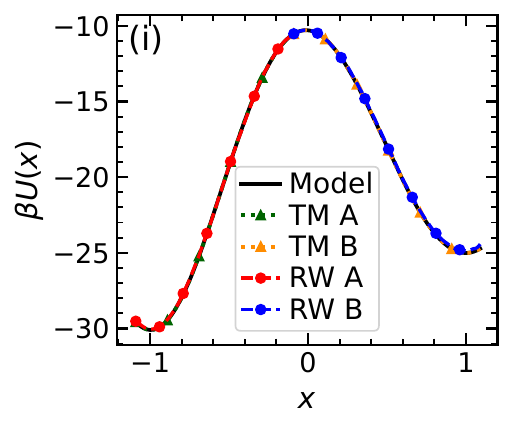}
        \label{fig:two_high_reconst}
    \end{subfigure}

    \caption[Two-region reconstruction for the low-, medium-, and high-barrier cases]
    {Two-region reconstruction for the low-, medium-, and high-barrier model potentials. The rows correspond to increasing barrier height. From left to right, the columns show the region-specific steady-state distributions $P_{\mathrm{st}}^A(x)$ and $P_{\mathrm{st}}^B(x)$, the corresponding MFPTs $\tau_A(x)$ and $\tau_B(x)$, and the reconstructed free-energy profile $\beta U(x)$. All the RW results are compared with TM calculations. In the free-energy panels, the model potential is compared with reconstructions obtained from the RW and TM quantities for regions~A and~B.}
    
    \label{fig:two_region_all_panels}
\end{figure*}

Fig.~\ref{fig:two_region_all_panels} shows the results in a $3\times3$ layout. The rows correspond to the low-, medium-, and high-barrier cases, while the columns show the region-specific steady-state distributions, MFPTs, and reconstructed free-energy profiles. For all three barrier heights, the RW estimates of $P_{\mathrm{st}}^{A,B}(x)$ and $\tau_{A,B}(x)$ closely follow the corresponding TM results. Reconstructions obtained from both the RW and TM quantities reproduce the model potential across the two regions. For these model potentials, the region-specific MFPTs decrease as the barrier height increases, reflecting the steeper descent from the barrier region toward the absorbing boundaries.

In contrast with the conventional Reguera calculation, the two-region simulations do not require trajectories to cross the entire barrier. Each simulation samples only one side of the barrier, so sufficient region-specific statistics can still be obtained as the barrier height increases. The high-barrier case therefore remains accessible without relying on spontaneous barrier crossings. These results validate the region-specific reconstruction and establish the reference behavior for the top-start procedure.

For the profiles shown in Fig.~\ref{fig:two_region_all_panels}, the two regional reconstructions are independently aligned to the model potential at one interior reference point in each region. To test the procedure for matching profiles from region~A and region~B  described in Section~\ref{subsec:kinetic_reconstruct_method}, we apply it to the two-region reconstruction using only one reference energy in region~A. For all three barrier heights, the extrapolated value of $\beta U(a)$ from the RW data agrees with the model potential to within approximately $0.007$, while the corresponding TM deviations are below $0.002$. This indicates that the relative offset between the two regional profiles can be determined by boundary matching, leaving only a single overall free-energy reference to be specified.

\subsection{Top-start reconstruction}

Finally, we apply the top-start scheme introduced in Section~\ref{subsec:top_start}. Each walker is initialized at $a=-0.1$ and evolves until reaching either absorbing boundary, $b_1=-1.1$ or $b_2=1.1$. The walker is then reinjected at $a$, and the process is repeated until the opposite absorbing boundary is reached, completing one top-start trajectory. In the top-start scheme, $a$ is an initial position rather than a reflecting boundary, and the absorbing boundary reached first may be visited multiple times before termination. 
The region-specific steady-state distributions and MFPTs are extracted from the top-start trajectories using the decomposition described in Appendix~\ref{app:top_start_RW_example}.

Fig.~\ref{fig:top_start_all_panels} uses the same $3\times3$ layout as Fig.~\ref{fig:two_region_all_panels} to facilitate comparison between the two methods.
For all three barrier heights, the $P_{\mathrm{st}}^{A,B}(x)$ and $\tau_{A,B}(x)$ obtained from the top-start RW trajectories closely follow the TM benchmarks. The free-energy profiles reconstructed from the RW data reproduce the model potential across both regions, including for the high-barrier case.

The top-start scheme preserves the sampling advantage of the two-region construction while removing the need for two separate reflecting-boundary simulations. Sufficient region-specific statistics can still be obtained without requiring full barrier crossings. The TM reconstruction is omitted because it has already been shown in Fig.~\ref{fig:two_region_all_panels} and closely reproduces the model potential.
\begin{figure*}[t]
    \centering
    \begin{subfigure}[b]{0.32\textwidth}
        \centering
        \includegraphics[width=\linewidth]{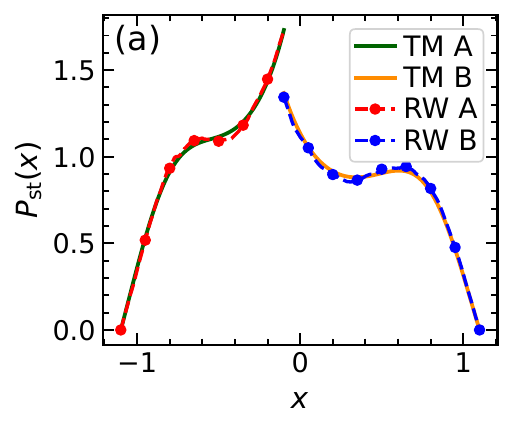}
        \label{fig:top_low_pst}
    \end{subfigure}\hfill
    \begin{subfigure}[b]{0.32\textwidth}
        \centering
        \includegraphics[width=\linewidth]{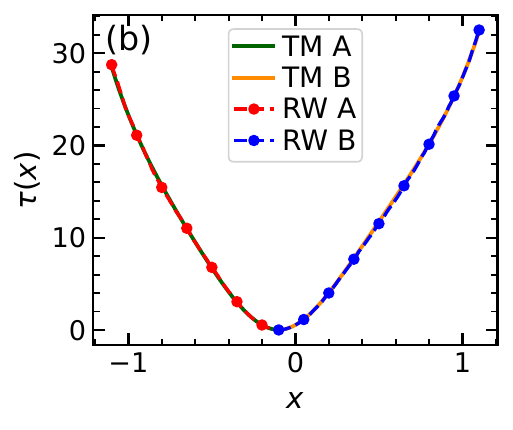}
        \label{fig:top_low_mfpt}
    \end{subfigure}\hfill
    \begin{subfigure}[b]{0.32\textwidth}
        \centering
        \includegraphics[width=\linewidth]{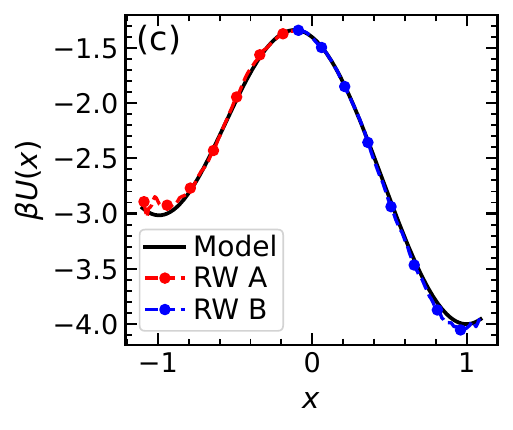}
        \label{fig:top_low_reconst}
    \end{subfigure}

    \begin{subfigure}[b]{0.32\textwidth}
        \centering
        \includegraphics[width=\linewidth]{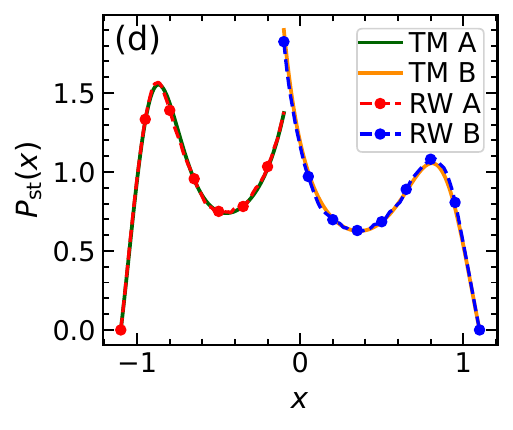}
        \label{fig:top_medium_pst}
    \end{subfigure}\hfill
    \begin{subfigure}[b]{0.32\textwidth}
        \centering
        \includegraphics[width=\linewidth]{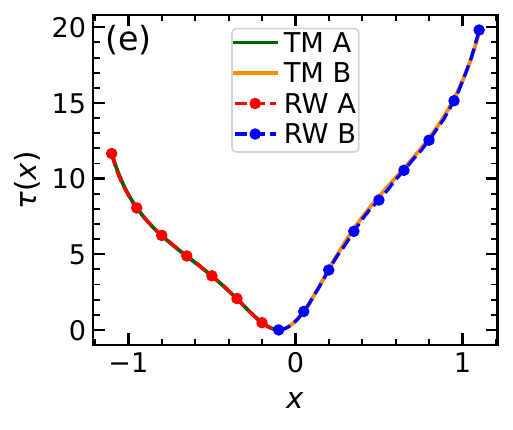}
        \label{fig:top_medium_mfpt}
    \end{subfigure}\hfill
    \begin{subfigure}[b]{0.32\textwidth}
        \centering
        \includegraphics[width=\linewidth]{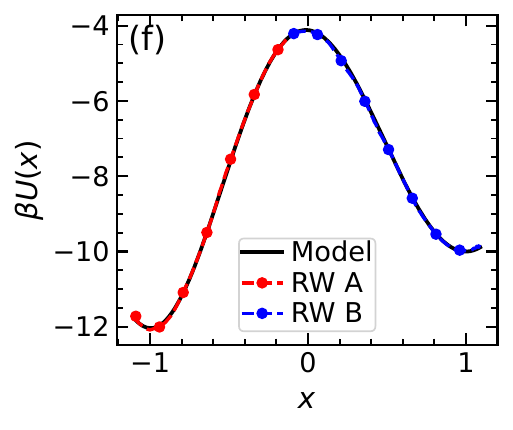}
        \label{fig:top_medium_reconst}
    \end{subfigure}

    \begin{subfigure}[b]{0.32\textwidth}
        \centering
        \includegraphics[width=\linewidth]{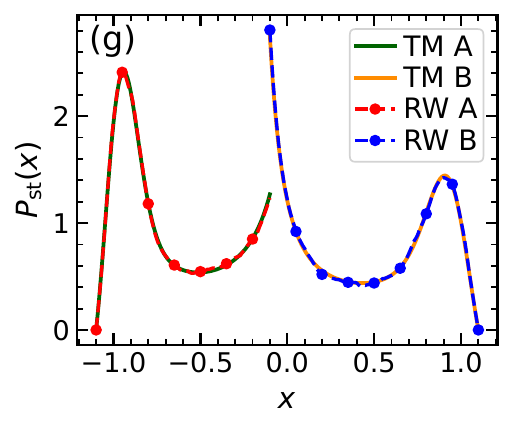}
        \label{fig:top_high_pst}
    \end{subfigure}\hfill
    \begin{subfigure}[b]{0.32\textwidth}
        \centering
        \includegraphics[width=\linewidth]{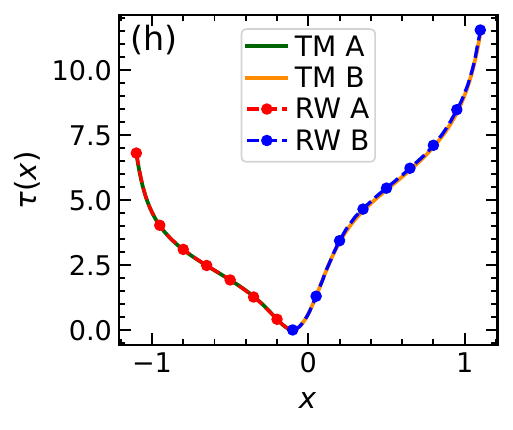}
        \label{fig:top_high_mfpt}
    \end{subfigure}\hfill
    \begin{subfigure}[b]{0.32\textwidth}
        \centering
        \includegraphics[width=\linewidth]{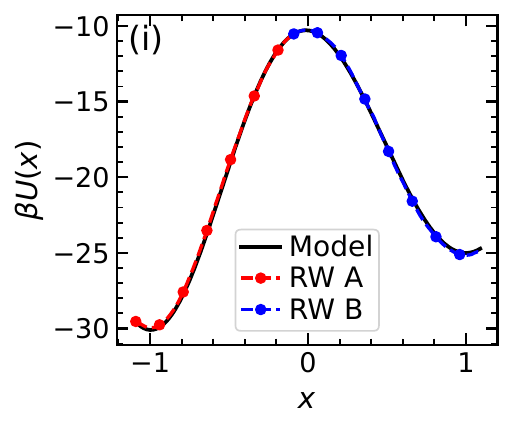}
        \label{fig:top_high_reconst}
    \end{subfigure}
    
    \caption[Top-start reconstruction for the low-, medium-, and high-barrier cases]
    {Top-start reconstruction for the low-, medium-, and high-barrier model potentials. The rows correspond to increasing barrier height. From left to right, the columns show the region-specific steady-state distributions $P_{\mathrm{st}}^A(x)$ and $P_{\mathrm{st}}^B(x)$, the corresponding MFPTs $\tau_A(x)$ and $\tau_B(x)$, and the reconstructed free-energy profile $\beta U(x)$. The steady-state distributions and MFPTs obtained from the top-start RW trajectories are compared with TM calculations, while the reconstructed free-energy profiles are compared directly with the corresponding model potentials.}
    \label{fig:top_start_all_panels}
\end{figure*}

\clearpage

\section{\label{sec:final}Discussion and Conclusion}

In this paper, we extend Reguera's MFPT-based free-energy reconstruction formalism to systems with high free-energy barriers through the development of a top-start scheme. The method builds on a two-region construction in which the steady-state distributions and MFPTs are obtained separately on the two sides of the barrier. 
In the top-start scheme, trajectories are instead initiated near the top of the barrier and decomposed into region-restricted trajectories to obtain the corresponding region-specific statistics, avoiding the need for separate reflecting-boundary simulations.

The equivalence between the top-start and two-region constructions is established using a transition-matrix description and further validated with MC random-walk simulations for low-, medium-, and high-barrier double-well potentials. For all three barrier heights, the RW estimates of the region-specific steady-state distributions and MFPTs closely follow the TM results, and the reconstructed free-energy profiles reproduce the model potentials. 
In contrast with the conventional Reguera reconstruction, the region-based reconstructions obtain the required statistics from trajectories initialized near the barrier top, avoiding the rapidly increasing cost of sampling full barrier crossings at high barriers.

For simulations of more complex systems, the preparation of starting configurations near the transition state is a necessary step.  For crystal nucleation, for example, this may be accomplished by using MC simulations with a biasing potential to harvest a set of configurations containing a near-critical nucleus.  We plan to test the top-start method for nucleation in simple liquids in future work.

Overall, the top-start scheme provides a  computationally efficient extension of Reguera's MFPT reconstruction for rare-event systems for which configurations can be prepared near the barrier top. The method therefore offers a route to reconstructing free-energy profiles from non-equilibrium trajectory data when direct sampling of spontaneous barrier crossings is prohibitively expensive.

\begin{acknowledgments}
I.S.-V. wishes to acknowledge the Natural Sciences and Engineering Research Council (NSERC) for funding support (grant number RGPIN-2023-05916).
\end{acknowledgments}

\section*{Data Availability Statement}

The data that support the findings of this study are openly available in a GitHub repository~\cite{MFPT_top_start_github}.

\appendix

\section{\label{sec:appendix_A}Basic concepts of finite Markov chains}

This appendix summarizes the finite-Markov-chain results used in the main text, following the classical treatment of Kemeny and Snell~\cite{kemeny1960finite}. We review absorbing and regular Markov chains and the relations between their transition matrices, steady-state distributions, and MFPTs.

A finite Markov chain is a stochastic process on a finite state space that satisfies the Markov property: the probability of the next state depends only on the current state and not on the earlier history of the process. All Markov chains considered in this work are time-homogeneous, so the one-step transition probability from state $s_i$ to state $s_j$ is independent of the time step and is denoted by $P_{ij}$.

The corresponding transition matrix is,
\begin{equation}
P=\{P_{ij}\},
\end{equation}
where row $i$ contains the probabilities of all possible transitions out of state $s_i$, so that
\begin{equation}
\sum_j P_{ij}=1.
\end{equation}
Probability distributions are represented as row vectors throughout this work. If $\boldsymbol{\rho}^{\,0}$ denotes the initial probability distribution, then the probability distribution after $n$ steps is,
\begin{equation}
\boldsymbol{\rho}^{\,n}=\boldsymbol{\rho}^{\,0}P^n.
\end{equation}

The matrices $P$, $P_A$, and $P_B$ introduced in Eqs.~\eqref{eq:AbAb_matrix}, \eqref{eq:AbRe_matrix}, and \eqref{eq:ReAb_matrix} describe absorbing chains. Replacing the absorbing boundaries by the recycling transitions introduced in the main text gives the corresponding recycling chains, with transition matrices $P'$, $P_A'$, and $P_B'$ defined in Eqs.~\eqref{eq:recycle_matrix}, \eqref{eq:recycle_A_matrix}, and~\eqref{eq:recycle_B_matrix}.

\subsection{\label{subsec:absorbing_chain_def}Absorbing Markov chains}

A state $s_i$ is \textit{absorbing} if, once entered, it cannot be left. In terms of the transition matrix, $P_{ii}=1$ and therefore $P_{ij}=0$ for all $j\neq i$. A finite Markov chain is an \textit{absorbing Markov chain} if it contains at least one absorbing state and every state can eventually reach an absorbing state~\cite{kemeny1960finite}. All nonabsorbing states are then transient.

\textbf{Theorem A.1.1.}
\textit{For an absorbing Markov chain, the probability that the process is in an absorbing state tends to $1$ as $n\to\infty$, regardless of the initial state.}

After reordering the states so that the absorbing states appear first and the transient states follow, the transition matrix can be written in canonical form as,
\begin{equation}
P=
\begin{bmatrix}
I & O\\
R & Q
\end{bmatrix},
\label{eq:absorbing_canonical_form}
\end{equation}
where $I$ is the identity matrix associated with the absorbing states, $O$ is a zero matrix, $Q$ contains transitions among the transient states, and $R$ contains transitions from transient to absorbing states. For an absorbing chain, $Q^n\rightarrow 0$ as $n\rightarrow\infty$.

The corresponding fundamental matrix is,
\begin{equation}
N=(I-Q)^{-1}
  =I+Q+Q^2+\cdots
  =\sum_{k=0}^{\infty}Q^k.
\label{eq:absorbing_fundamental_matrix}
\end{equation}

\textbf{Theorem A.1.2.}
\textit{For an absorbing Markov chain, the $(i,j)$ entry of the fundamental matrix $N$ gives the expected number of visits to transient state $s_j$ before absorption when the process starts from transient state $s_i$.}

Thus, each row of $N$ gives the expected pre-absorption occupation counts over the transient states for a specified starting state. This result provides the connection between the transition-matrix formulation and the visit counts used in the reconstruction procedure; the relation to the steady-state distribution of the corresponding recycling process is established in Appendix~\ref{sec:appendix_B}.

\textbf{Theorem A.1.3.}
\textit{Let $b_{ij}$ denote the probability that a process starting from transient state $s_i$ is eventually absorbed in absorbing state $s_j$. The matrix of absorption probabilities is}
\begin{equation}
B=NR=(I-Q)^{-1}R.
\label{eq:absorption_probability_matrix}
\end{equation}

For each transient starting state, the corresponding row of $B$ therefore gives the probabilities of eventual absorption in the available absorbing states, and these probabilities sum to unity.

\subsection{\label{subsec:regular_chain_def}Regular Markov chains}

The recycling chains introduced in the main text, with transition matrices $P'$, $P_A'$, and $P_B'$, are regular for the systems considered here. The properties of regular Markov chains therefore provide the steady-state distributions and MFPTs used in the transition-matrix calculations.

A finite Markov chain is \textit{regular} if there exists a positive integer $k$ such that every entry of $P^k$ is strictly positive~\cite{Sekhon2022Regular, kemeny1960finite}. Equivalently, after the same finite number of steps, every state can be reached from every starting state with nonzero probability.

\textbf{Theorem A.2.1.}
\textit{Let $P$ be the transition matrix of a regular Markov chain. Then, for any initial probability distribution $\boldsymbol{\rho}^{\,0}$,
\begin{equation}
\lim_{n\to\infty}\boldsymbol{\rho}^{\,0}P^n
=\boldsymbol{\alpha},
\end{equation}
where $\boldsymbol{\alpha}$ is independent of the initial distribution, has strictly positive components, and satisfies,
\begin{equation}
\boldsymbol{\alpha}P=\boldsymbol{\alpha},
\qquad
\sum_i\alpha_i=1.
\end{equation}
Furthermore,
\begin{equation}
\lim_{n\to\infty}P^n=A,
\end{equation}
where every row of the limiting matrix $A$ is equal to $\boldsymbol{\alpha}$, so that,
\begin{equation}
A=\boldsymbol{\xi}\boldsymbol{\alpha},
\end{equation}
where $\boldsymbol{\xi}$ is a column vector of ones.}

The matrix $A$ is the limiting matrix of the regular chain, and the vector $\boldsymbol{\alpha}$ is its limiting probability distribution. For consistency with the terminology used in the reconstruction formalism, we refer to $\boldsymbol{\alpha}$ as the steady-state distribution of the regular chain. The relation between this distribution and the normalized pre-absorption occupation counts of the corresponding absorbing process is established in Appendix~\ref{sec:appendix_B}.

For a regular Markov chain, the fundamental matrix is defined by the following result.

\textbf{Theorem A.2.2.}
\textit{Let $P$ be the transition matrix for a regular Markov chain, and let $A$ be the limiting matrix associated with $P$. Then the fundamental matrix,
\begin{equation}
Z=\left[I-(P-A)\right]^{-1},
\label{eq:regular_fundamental_matrix}
\end{equation}
exists and satisfies the relation,
\begin{equation}
Z=I+\sum_{n=1}^{\infty}(P^n-A).
\end{equation}}

The fundamental matrix $Z$ can be used to obtain the MFPTs between states of the regular chain. Let $M_{ij}$ denote the mean first-passage time from state $s_i$ to state $s_j$, and let $M=\{M_{ij}\}$ denote the corresponding mean first-passage matrix.

\textbf{Theorem A.2.3.}
\textit{The mean first-passage matrix is,
\begin{equation}
M=(I-Z+EZ_d)D,
\label{eq:mean_first_passage_matrix}
\end{equation}
where $E$ is the matrix of ones, $Z_d$ is the diagonal matrix containing the diagonal elements of $Z$, and $D$ is the diagonal matrix with elements,
\begin{equation}
D_{ii}=\frac{1}{\alpha_i},
\end{equation}
where $\alpha_i$ is the $i$-th component of the steady-state distribution $\boldsymbol{\alpha}$.}

For $i\neq j$, the entry $M_{ij}$ is the MFPT from state $s_i$ to state $s_j$. The diagonal entry $M_{ii}=1/\alpha_i$ instead represents the mean first-return time to $s_i$. In the reconstruction formalism, the first-passage time from a state to itself is taken to be zero. We therefore define,
\begin{equation}
\bar{M}=M-D,
\label{eq:modified_mfpt_matrix}
\end{equation}
which leaves all off-diagonal MFPTs unchanged and sets the diagonal entries to zero. The matrix $\bar{M}$ is used to obtain the MFPT between any chosen starting and target states from the transition matrix of the corresponding regular chain.

\section{\label{sec:appendix_B}Applications of finite-Markov-chain theory}

This appendix applies the results summarized in Appendix~\ref{sec:appendix_A} to establish the relation between the steady-state distribution and pre-absorption occupation counts, and to validate the top-start scheme.

\subsection{\label{subsec:prove_steady_state}Relation between the steady-state distribution and pre-absorption occupation counts}

Consider an absorbing Markov chain with a single absorbing state. After placing the absorbing state first, its transition matrix has the canonical form,
\begin{equation}
P=
\begin{bmatrix}
1 & \boldsymbol{0}\\
R & Q
\end{bmatrix},
\end{equation}
where $Q$ contains transitions among the transient states, $R$ contains transitions from the transient states to the absorbing state, and $\boldsymbol{0}$ is a zero row vector.

Let the probability vector,
\[
\boldsymbol{\eta}=(\eta_1,\boldsymbol{\eta}_2),
\]
denote the reinjection distribution, where $\eta_1$ corresponds to the absorbing state and $\boldsymbol{\eta}_2$ to the transient states. In the processes considered in this work, the same distribution is used to initialize the process and to reinject it after absorption. Replacing the absorbing row of the canonical transition matrix by $\boldsymbol{\eta}$ gives the recycling transition matrix,
\begin{equation}
P'=
\begin{bmatrix}
\eta_1 & \boldsymbol{\eta}_2\\
R & Q
\end{bmatrix}.
\end{equation}

Let the steady-state distribution of the recycling chain be partitioned as,
\[
\boldsymbol{\alpha}=(\alpha_1,\boldsymbol{\alpha}_2),
\]
where $\alpha_1$ corresponds to the state that was absorbing in the original absorbing chain, and $\boldsymbol{\alpha}_2$ contains the components associated with its transient states. From Theorem~A.2.1,
\[
\boldsymbol{\alpha}P'=\boldsymbol{\alpha},
\]
with strictly positive components of $\boldsymbol{\alpha}$. Evaluating this relation blockwise gives,
\begin{align}
\alpha_1\eta_1+\boldsymbol{\alpha}_2R&=\alpha_1,
\label{eq:alpha1}\\
\alpha_1\boldsymbol{\eta}_2+\boldsymbol{\alpha}_2Q
&=\boldsymbol{\alpha}_2.
\label{eq:alpha2}
\end{align}
Since $\alpha_1>0$, Eq.~\eqref{eq:alpha2} gives,
\begin{equation}
\frac{\boldsymbol{\alpha}_2}{\alpha_1}
=
\boldsymbol{\eta}_2(I-Q)^{-1}
=
\boldsymbol{\eta}_2N,
\label{eq:alpha2_bar}
\end{equation}
where $N=(I-Q)^{-1}$ is the fundamental matrix of the corresponding absorbing chain.

According to Theorem~A.1.2, $\boldsymbol{\eta}_2N$ gives the expected pre-absorption occupation counts of the transient states for the initial distribution $\boldsymbol{\eta}_2$. 
In the processes considered in this work, reinjection occurs at a specified transient starting state, so $\eta_1=0$ and $\boldsymbol{\eta}_2$ contains a single nonzero entry equal to unity. Therefore, $\boldsymbol{\eta}_2N$ is the corresponding row of $N$.

Eq.~\eqref{eq:alpha2_bar} shows that $\boldsymbol{\alpha}_2$ (the vector portion of the steady-state distribution) is proportional to the expected pre-absorption occupation counts $\boldsymbol{\eta}_2N$. Therefore,
\begin{equation}
\frac{\boldsymbol{\alpha}_2}
{\boldsymbol{\alpha}_2\boldsymbol{\xi}}
=
\frac{\boldsymbol{\eta}_2N}
{\boldsymbol{\eta}_2N\boldsymbol{\xi}},
\label{eq:normalized_occupation_equivalence}
\end{equation}
where $\boldsymbol{\xi}$ is a column vector of ones of the appropriate dimension. The normalized distributions are therefore identical, allowing the steady-state distribution used in the reconstruction formalism to be obtained from pre-absorption occupation counts without waiting for the recycling process to converge to its limiting probability distribution. 
Thus, a normalized visit-count histogram obtained from an ensemble of trajectories starting at the initial state and ending at the absorbing state provides a rigorous estimate of the corresponding steady-state distribution of the recycling process.

The remaining relation, Eq.~\eqref{eq:alpha1}, gives, after division by $\alpha_1$ and use of Eq.~\eqref{eq:alpha2_bar},
\begin{equation}
\eta_1+\boldsymbol{\eta}_2NR=1.
\end{equation}
According to Theorem~A.1.3, $NR$ gives the probabilities of eventual absorption from the transient states into the original absorbing state. The first term therefore accounts for reinjection directly into the absorbing state, whereas the second accounts for eventual absorption following reinjection into the original transient states. For the processes considered here, $\eta_1=0$ and there is only one absorbing state, so $\boldsymbol{\eta}_2NR=1$, consistent with Theorem~A.1.1.

\subsection{\label{subsec:prove_top_start}Transition-matrix equivalence of the top-start and two-region constructions}

We now consider the recycling chain associated with the top-start scheme. Whenever the process reaches either boundary state, $s_0$ or $s_{m+1}$, it is reinjected at the starting state $s_r$. For the finite systems considered here, this recycling chain is regular and therefore has a unique steady-state distribution. The termination rule used to define a finite top-start realization determines when the realization ends, but does not alter the transition probabilities. It is therefore not needed in deriving the region-restricted dynamics.

We first consider a general Markov chain with transition matrix $P$, and suppose that it is observed only when it lies in a subset $\mathcal{S}$, which we call the observable set, with complement $\widetilde{\mathcal{S}}$. Retaining only the sequence of visits to $\mathcal{S}$ defines a reduced Markov chain on $\mathcal{S}$. A transition of the reduced chain begins at a state in $\mathcal{S}$ and ends when the full chain is next observed in $\mathcal{S}$; any intermediate states in $\widetilde{\mathcal{S}}$ are omitted.

After relabeling the states so that those in $\mathcal{S}$ appear first, the transition matrix can be written in block form as
\begin{equation}
P=
\begin{bmatrix}
T & U\\
R & Q
\end{bmatrix},
\end{equation}
where $T$ contains transitions within $\mathcal{S}$, $U$ transitions from $\mathcal{S}$ to $\widetilde{\mathcal{S}}$, $R$ transitions from $\widetilde{\mathcal{S}}$ back to $\mathcal{S}$, and $Q$ transitions within $\widetilde{\mathcal{S}}$.

A transition of the reduced chain occurs either directly within $\mathcal{S}$ through $T$, or after an excursion through $\widetilde{\mathcal{S}}$. The latter contribution is,
\[
U\left(I+Q+Q^2+\cdots\right)R
=
U(I-Q)^{-1}R.
\]
Therefore, the transition matrix of the reduced chain is,
\begin{equation}
\bar{P}=T+U(I-Q)^{-1}R.
\label{eq:reduced_transition_matrix}
\end{equation}

A useful consequence of this construction concerns the steady-state distribution. Let the steady-state distribution of $P$ be partitioned as,
\[
\boldsymbol{\alpha}
=
(\boldsymbol{\alpha}_{\mathcal{S}},
\boldsymbol{\alpha}_{\widetilde{\mathcal{S}}}),
\]
where $\boldsymbol{\alpha}_{\mathcal{S}}$ contains the components associated with $\mathcal{S}$ and $\boldsymbol{\alpha}_{\widetilde{\mathcal{S}}}$ contains those associated with $\widetilde{\mathcal{S}}$. Evaluating $\boldsymbol{\alpha}P=\boldsymbol{\alpha}$ blockwise gives,
\begin{align}
\boldsymbol{\alpha}_{\mathcal{S}}
&=
\boldsymbol{\alpha}_{\mathcal{S}}T
+
\boldsymbol{\alpha}_{\widetilde{\mathcal{S}}}R,
\label{eq:reduced_alpha_S}
\\
\boldsymbol{\alpha}_{\widetilde{\mathcal{S}}}
&=
\boldsymbol{\alpha}_{\mathcal{S}}U
+
\boldsymbol{\alpha}_{\widetilde{\mathcal{S}}}Q.
\label{eq:reduced_alpha_complement}
\end{align}
Eq.~\eqref{eq:reduced_alpha_complement} gives,
\[
\boldsymbol{\alpha}_{\widetilde{\mathcal{S}}}
=
\boldsymbol{\alpha}_{\mathcal{S}}
U(I-Q)^{-1}.
\]
Substituting this result into Eq.~\eqref{eq:reduced_alpha_S} yields,
\[
\boldsymbol{\alpha}_{\mathcal{S}}
=
\boldsymbol{\alpha}_{\mathcal{S}}
\left[T+U(I-Q)^{-1}R\right]
=
\boldsymbol{\alpha}_{\mathcal{S}}\bar{P}.
\]
Thus, after normalization over $\mathcal{S}$,
\begin{equation}
\bar{\boldsymbol{\alpha}}_{\mathcal{S}}
=
\frac{\boldsymbol{\alpha}_{\mathcal{S}}}
{\boldsymbol{\alpha}_{\mathcal{S}}\boldsymbol{\xi}},
\label{eq:reduced_steady_state}
\end{equation}
where $\boldsymbol{\xi}$ is a column vector of ones of the appropriate dimension. The normalized vector $\bar{\boldsymbol{\alpha}}_{\mathcal{S}}$ is therefore the steady-state distribution of the reduced chain. Hence, the steady-state distribution of the reduced chain can be obtained directly by restricting the steady-state distribution of the full chain to $\mathcal{S}$ and renormalizing over that set.

We now apply this construction to the recycling chain associated with the top-start scheme, whose transition matrix $P'$ is defined in Eq.~\eqref{eq:recycle_matrix}. For region~A, the observable set is
\[
\mathcal{S}_A=\{s_0,s_1,\ldots,s_r\},
\]
with complement,
\[
\widetilde{\mathcal{S}}_A
=
\{s_{r+1},\ldots,s_{m+1}\}.
\]
The corresponding block $T$ is,
\begin{equation}
T=
\begin{bmatrix}
       &        &        &        &        & 1      \\
\ell_1    & h_1    & r_1    &        &        &        \\
       & \ell_2    & h_2    & r_2    &        &        \\
       &        & \ddots & \ddots & \ddots &        \\
       &        &        & \ell_{r-1}& h_{r-1}& r_{r-1}\\
       &        &        &        & \ell_r    & h_r
\end{bmatrix},
\label{eq:submatrix_T}
\end{equation}
while $U$ contains only one nonzero element, $r_r$, corresponding to the transition from $s_r$ to $s_{r+1}$.

To evaluate the excursion contribution in Eq.~\eqref{eq:reduced_transition_matrix}, we temporarily treat the states in $\mathcal{S}_A$ as absorbing. According to Theorem~A.1.3, the matrix,
\begin{equation}
B=(I-Q)^{-1}R,
\label{eq:matrixB_reduced}
\end{equation}
gives the probabilities that, starting from any given state in $\widetilde{\mathcal{S}}_A$, the chain first enters $\mathcal{S}_A$ through each of its states. For the present Markov chain with transition matrix $P'$, any return from $\widetilde{\mathcal{S}}_A$ to $\mathcal{S}_A$ must occur at $s_r$. The process either returns through the transition $s_{r+1}\rightarrow s_r$ or reaches $s_{m+1}$ and is reinjected at $s_r$. Hence,
\begin{equation}
B=
\begin{bmatrix}
0 & \cdots & 0 & 1\\
0 & \cdots & 0 & 1\\
\vdots & & \vdots & \vdots\\
0 & \cdots & 0 & 1
\end{bmatrix},
\label{eq:matrixB_setS}
\end{equation}
where the final column corresponds to $s_r$.

Since $U$ has only the single nonzero element $r_r$, the product $UB$ has only one nonzero element, also equal to $r_r$, in the entry corresponding to $s_r\rightarrow s_r$. Eq.~\eqref{eq:reduced_transition_matrix} therefore gives,
\begin{equation}
\bar{P}_A=
\begin{bmatrix}
       &        &        &        &        & 1      \\
\ell_1    & h_1    & r_1    &        &        &        \\
       & \ell_2    & h_2    & r_2    &        &        \\
       &        & \ddots & \ddots & \ddots &        \\
       &        &        & \ell_{r-1}& h_{r-1}& r_{r-1}\\
       &        &        &        & \ell_r    & h_r+r_r
\end{bmatrix}
=
P_A',
\label{eq:reduced_matrix_A}
\end{equation}
where $P_A'$ is defined in Eq.~\eqref{eq:recycle_A_matrix}. Thus, the transition matrix of the reduced chain obtained by retaining only region-A states is identical to $P_A'$, which corresponds to the region-A recycling chain.

The same construction applies to region~B, for which,
\[
\mathcal{S}_B=\{s_r,s_{r+1},\ldots,s_{m+1}\}.
\]
In this case, excursions through region~A are omitted, and the resulting reduced transition matrix satisfies,
\begin{equation}
\bar{P}_B=P_B',
\label{eq:reduced_matrix_B}
\end{equation}
where $P_B'$ is defined in Eq.~\eqref{eq:recycle_B_matrix}.

Equations~\eqref{eq:reduced_matrix_A} and~\eqref{eq:reduced_matrix_B} show that the reduced transition matrices obtained from the top-start scheme are identical to the corresponding region-specific recycling transition matrices, $P_A'$ and $P_B'$. The corresponding Markov chains are therefore equivalent and give identical steady-state distributions and MFPTs. This establishes the transition-matrix equivalence underlying the top-start reconstruction.

\section{\label{app:top_start_RW_example}Illustration of the top-start sampling procedure}

\begin{figure*}[htbp]
    \centering
    \includegraphics[width=0.48\textwidth]{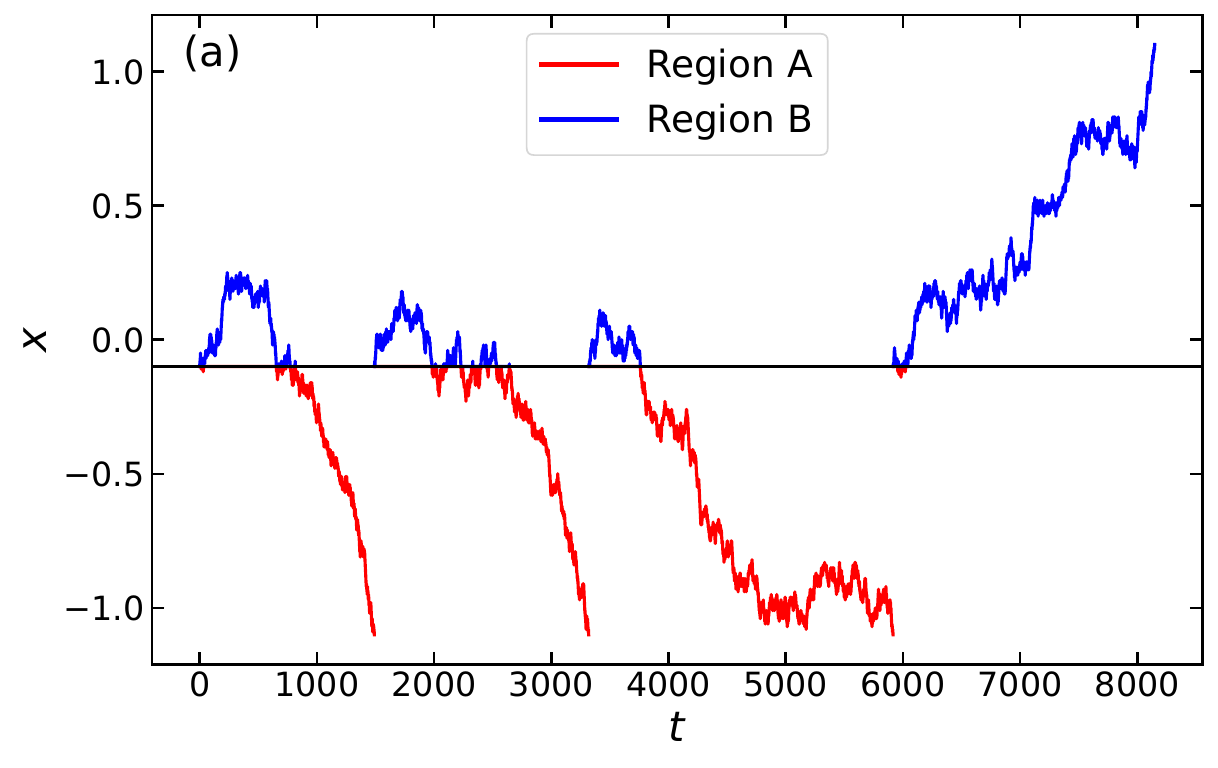}
    \includegraphics[width=0.48\textwidth]{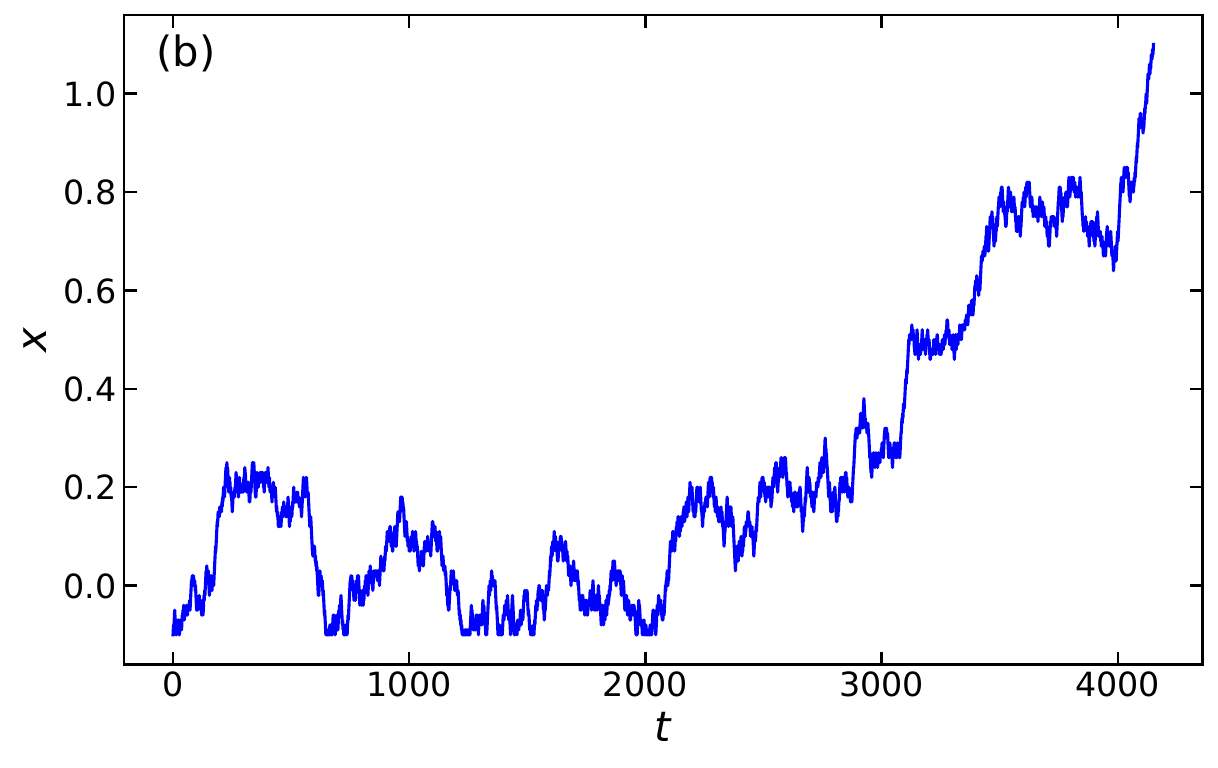}\\[2mm]
    \includegraphics[width=0.27\textwidth]{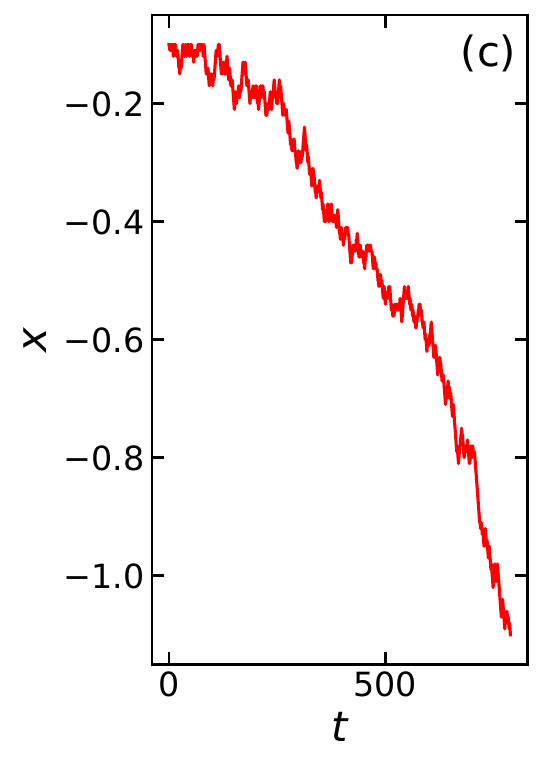}
    \includegraphics[width=0.27\textwidth]{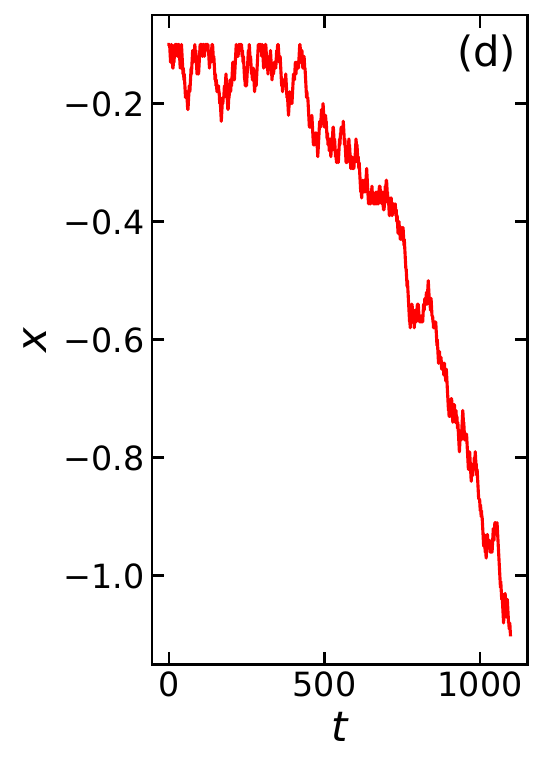}
    \includegraphics[width=0.35\textwidth]{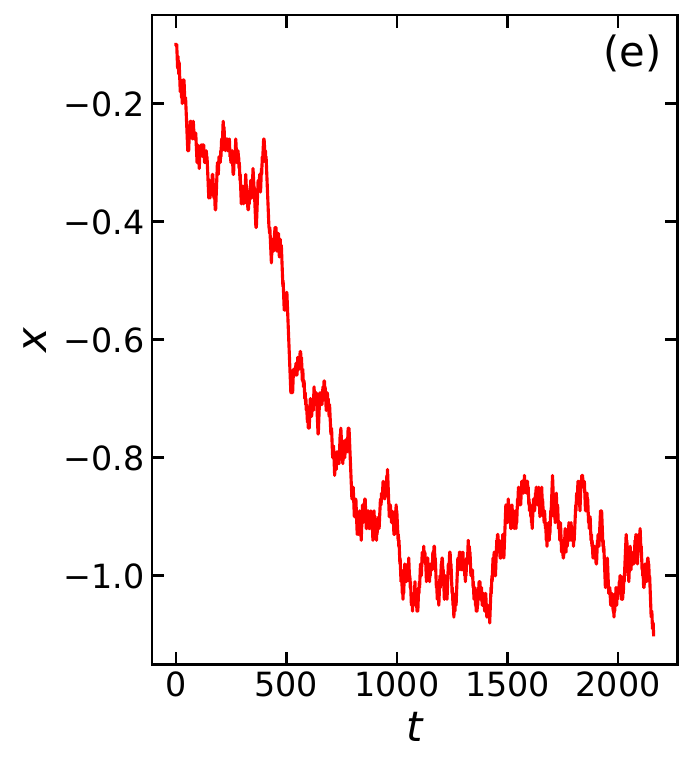}
    \caption[Example decomposition of a top-start random-walk trajectory]{
    Example decomposition of a representative top-start random-walk trajectory.
    (a) Full top-start trajectory starting from $a = -0.1$ and evolving with recycling at the absorbing boundaries.
    (b) A completed region-B trajectory obtained by retaining the states visited in region~B and omitting excursions into region~A.
    (c)--(e) Three completed region-A trajectories obtained analogously.
    A fourth region-A trajectory begins before the top-start realization terminates but remains incomplete and is discarded.
    The region-restricted time of each trajectory excludes the omitted excursions.}
    \label{fig:top_start_example}
\end{figure*}

We first illustrate the trajectory decomposition using a representative top-start realization obtained from the random-walk simulations. For this example, the starting position is $a=-0.1$, with absorbing boundaries at $b_1=-1.1$ for region~A and $b_2=1.1$ for region~B. Fig.~\ref{fig:top_start_example} shows the full top-start trajectory together with the completed region-restricted trajectories extracted from it. Panel~(a) shows the full top-start trajectory, panel~(b) shows one completed region-B trajectory, and panels~(c)--(e) show three completed region-A trajectories. For each region, only states in that region are retained, while excursions into the opposite region are omitted; the corresponding region-restricted time advances only along the retained trajectory. After the third region-A trajectory is completed, a fourth region-A trajectory begins but remains incomplete when the top-start realization terminates; it is therefore discarded and is not shown separately.
The FPT arrays and visit-count histograms of the three completed region-A trajectories shown in panels~(c)--(e) are averaged position by position. Thus, the single top-start realization illustrated in Fig.~\ref{fig:top_start_example} contributes one FPT array and one visit-count histogram for each region. Averaging these quantities over independent top-start realizations yields $\tau_A(x)$, $P_{\mathrm{st}}^A(x)$, $\tau_B(x)$, and $P_{\mathrm{st}}^B(x)$.

To show the top-start procedure step by step, we next consider a six-state random-walk example with accessible states $x=\{-2,-1,0,1,2,3\}$. The absorbing boundaries are $b_1=-2$ and $b_2=3$, and the starting point is $a=0$. Region~A spans $b_1\leq x\leq a$, while region~B spans $a\leq x\leq b_2$. Whenever an absorbing boundary is reached before the top-start realization terminates, the walker is reinjected at $a$. The realization terminates once both $b_1$ and $b_2$ have been visited.

Table~\ref{tab:rw_full_table} gives one such full top-start trajectory. The walker reaches $b_1$ at $t=5$ and again at $t=14$, with reinjection at $a$ after each event. It reaches $b_2$ for the first time at $t=22$, at which point both absorbing boundaries have been visited and the top-start realization terminates.
\begin{table}[htbp]
\centering
\captionsetup{skip=8pt}
\caption[Example time series of a complete top-start random-walk realization]{
Example time series of a complete top-start random-walk realization.
The walker starts at $a=0$, is reinjected at $a$ after reaching an absorbing boundary, and the realization terminates once both $b_1=-2$ and $b_2=3$ have been visited.}
\label{tab:rw_full_table}
\begin{tabular}{c | c | l}
\hline
\textbf{Time step $t$} & \textbf{Position $x$} & \textbf{Region/event} \\
\hline
0   & 0   & Regions A and B (start at $a$) \\
1   & 1   & Region B \\
2   & 0   & Regions A and B (at $a$) \\
3   & -1  & Region A \\
4   & -1  & Region A \\
5   & -2  & Region A (reached $b_1$) \\
6   & 0   & Regions A and B (reinjected at $a$) \\
7   & 1   & Region B \\
8   & 2   & Region B \\
9   & 2   & Region B \\
10  & 1   & Region B \\
11  & 0   & Regions A and B (at $a$) \\
12  & 0   & Regions A and B (at $a$) \\
13  & -1  & Region A \\
14  & -2  & Region A (reached $b_1$) \\
15  & 0   & Regions A and B (reinjected at $a$) \\
16  & 0   & Regions A and B (at $a$) \\
17  & -1  & Region A \\
18  & 0   & Regions A and B (at $a$) \\
19  & 1   & Region B \\
20  & 1   & Region B \\
21  & 2   & Region B \\
22  & 3   & Region B (reached $b_2$; realization ends) \\
\hline
\end{tabular}
\end{table}
The full top-start trajectory generates region-restricted trajectories for regions~A and~B simultaneously. For region~A, states visited outside of  region~A are omitted, and similarly, for region~B, states visited outside region~B are omitted. The starting point $a$ belongs to both regions and is retained whenever it is visited. Importantly, reaching an absorbing boundary completes only the current region-restricted trajectory for that side, while the trajectory in the other region is not reset and continues after reinjection at $a$.

For example, the first region-A trajectory retains the original times $t=0,2,3,4,5$; the excursion to $x=1$ at $t=1$ is omitted. This trajectory reaches $b_1$ and is therefore complete. After reinjection at $a$, a second region-A trajectory begins at $t=6$ and retains $t=6,11,12,13,14$, again omitting the intervening region-B excursion. These trajectories are shown in Tables~\ref{tab:rw_A_segment1} and~\ref{tab:rw_A_segment2}. Their region-restricted time indices advance only along the retained states.
\begin{table}[htbp]
\centering
\caption[First completed region-A trajectory]{
First completed region-A trajectory extracted from Table~\ref{tab:rw_full_table}.
States visited in the opposite region are omitted when constructing the region-restricted time.}
\label{tab:rw_A_segment1}
\begin{tabular}{c | c | c}
\hline
\textbf{Position $x$} & \textbf{Original time $t$} & \textbf{Region-restricted time} \\
\hline
0  & 0 & 0 \\
0  & 2 & 1 \\
-1 & 3 & 2 \\
-1 & 4 & 3 \\
-2 & 5 & 4 \\
\hline
\end{tabular}
\end{table}

\begin{table}[htbp]
\centering
\caption[Second completed region-A trajectory]{
Second completed region-A trajectory extracted from Table~\ref{tab:rw_full_table}.}
\label{tab:rw_A_segment2}
\begin{tabular}{c | c | c}
\hline
\textbf{Position $x$} & \textbf{Original time $t$} & \textbf{Region-restricted time} \\
\hline
0  & 6  & 0 \\
0  & 11 & 1 \\
0  & 12 & 2 \\
-1 & 13 & 3 \\
-2 & 14 & 4 \\
\hline
\end{tabular}
\end{table}

The region-B trajectory, by contrast, begins at $t=0$ and remains the same region-restricted trajectory until $b_2$ is reached at $t=22$. The two visits to $b_1$ and the associated region-A excursions are omitted and therefore do not contribute to the region-restricted time of region~B. The region-B trajectory is not reset by those events. Table~\ref{tab:rw_B_segment} shows the resulting region-B trajectory.
\begin{table}[htbp]
\centering
\caption[Completed region-B trajectory]{
Completed region-B trajectory extracted from Table~\ref{tab:rw_full_table}.
Region-A excursions are omitted and therefore do not contribute to the region-restricted time, while the trajectory continues after reinjection at $a$.}
\label{tab:rw_B_segment}
\begin{tabular}{c | c | c}
\hline
\textbf{Position $x$} & \textbf{Original time $t$} & \textbf{Region-restricted time} \\
\hline
0  & 0  & 0 \\
1  & 1  & 1 \\
0  & 2  & 2 \\
0  & 6  & 3 \\
1  & 7  & 4 \\
2  & 8  & 5 \\
2  & 9  & 6 \\
1  & 10 & 7 \\
0  & 11 & 8 \\
0  & 12 & 9 \\
0  & 15 & 10 \\
0  & 16 & 11 \\
0  & 18 & 12 \\
1  & 19 & 13 \\
1  & 20 & 14 \\
2  & 21 & 15 \\
3  & 22 & 16 \\
\hline
\end{tabular}
\end{table}

A third region-A trajectory begins after $b_1$ is reached for the second time and retains $t=15,16,17,18$. It has not reached $b_1$ when the top-start realization terminates upon reaching $b_2$ at $t=22$. This region-restricted trajectory is therefore incomplete and is discarded from both the first-passage and visit-count statistics.

For each completed region-restricted trajectory, the first-passage time (FPT) to a state is the first region-restricted time at which that state is reached, and the visit count records the number of visits to that state along the same trajectory. Table~\ref{tab:rw_fpt_dist_summary} summarizes these quantities for the three completed region-restricted trajectories in the six-state example.
\begin{table}[htbp]
\centering
\caption[FPTs and visit counts from the completed region-restricted trajectories]{
FPTs and visit counts obtained from the two completed region-A trajectories and one completed region-B trajectory extracted from the example top-start trajectory.}
\label{tab:rw_fpt_dist_summary}
\begin{tabular}{c | c | c | c}
\hline
\textbf{Trajectory} & \textbf{Position $x$} & \textbf{FPT} & \textbf{Visit count} \\
\hline
A1 & 0  & 0  & 2 \\
A1 & -1 & 2  & 2 \\
A1 & -2 & 4  & 1 \\
\hline
A2 & 0  & 0  & 3 \\
A2 & -1 & 3  & 1 \\
A2 & -2 & 4  & 1 \\
\hline
B1 & 0  & 0  & 8 \\
B1 & 1  & 1  & 5 \\
B1 & 2  & 5  & 3 \\
B1 & 3  & 16 & 1 \\
\hline
\end{tabular}
\end{table}

In this top-start realization, two region-A trajectories are completed before the region-B trajectory is completed. To prevent a realization with multiple completions on one side from receiving greater weight for that side, the completed region-restricted trajectories for that side are first averaged within the realization. 
For the two completed region-A trajectories in Table~\ref{tab:rw_fpt_dist_summary}, the resulting FPTs at $x=(0,-1,-2)$ are $(0,2.5,4)$, and the corresponding visit counts are $(2.5,1.5,1)$.
Thus, each top-start realization contributes one region-specific FPT array and one visit-count histogram for each region.



Repeating the top-start procedure independently generates an ensemble of top-start realizations, each contributing a region-specific FPT array and visit-count histogram. In the random-walk simulations reported in Section~\ref{sec:random_walk_examples}, we reiterate that when multiple trajectories for the same region are completed within one realization, their FPTs and visit counts are first averaged position by position so that each top-start realization provides equal statistical weight for a given region. The resulting region-specific FPT arrays and visit-count histograms are then averaged over the ensemble to obtain $\tau_A(x)$, $\tau_B(x)$, $P_{\mathrm{st}}^A(x)$, and $P_{\mathrm{st}}^B(x)$. These quantities are used in Eqs.~\eqref{eq:auxiliary_regionA} and~\eqref{eq:reconstruct_A} for region~A and Eqs.~\eqref{eq:auxiliary_regionB} and~\eqref{eq:reconstruct_B} for region~B.

As a robustness check, we also test an alternative procedure in which only the first completed region-restricted trajectory for each region is retained to form the region-specific data for each realization. After averaging these data over the ensemble of top-start realizations, the final results are essentially the same as those obtained using the within-realization averaging procedure for the random-walk simulations reported in Section~\ref{sec:random_walk_examples}.

\bibliography{references}

\end{document}